\documentclass{aa}
\usepackage{graphicx}
\usepackage{txfonts}
\usepackage{url}
\usepackage{natbib} 
\bibpunct{(}{)}{;}{a}{}{,} % to follow the A&A style
\begin{document}
   \title{Effects of rotation on the evolution of primordial stars}

%   \subtitle{}

   \author{S. Ekstr\"om\inst{1},
         G. Meynet\inst{1},
         C. Chiappini\inst{1,2},
          R. Hirschi\inst{3}
          \and
          A. Maeder\inst{1}}

   \authorrunning{Ekstr\"om et al.}

   \offprints{Sylvia.Ekstrom@obs.unige.ch}

   \institute{Geneva Observatory, University of Geneva, Maillettes 51, 1290 Sauverny, Switzerland
          \and Osservatorio Astronomico di Trieste, Via G.B. Tiepolo 11, I - 34131 Trieste, Italia
          \and University of Keele, Keele, ST5 5BG, UK}

   \date{}

% \abstract{}{}{}{}{} 
% 5 {} token are mandatory
 
  \abstract
   {Though still beyond our observational abilities, Population III stars are interesting objects in many perspectives. They are responsible for the re-ionisation of the inter-galactic medium. They also left their chemical imprint in the early Universe, imprint which can be deciphered in the most metal-poor stars in the halo of our Galaxy.}
   {Rotation has been shown to play a determinant role at very low metallicity, bringing heavy mass loss where almost none was expected. Is this still true when the metallicity strictly equals zero? The aim of our study is to get an answer to this question, and to determine how rotation changes the evolution and the chemical signature of the primordial stars.}
   {We have calculated seven differentially-rotating stellar models at zero metallicity, with masses between 9 and 200 $M_{\sun}$. For each mass, we have also calculated a corresponding model without rotation. The evolution has been followed up to the pre-supernova stage.}
{We find that $Z=0$ models rotate with an internal profile $\Omega(r)$ close to local angular momentum conservation, because of a very weak core-envelope coupling. Rotational mixing drives a H-shell boost due to a sudden onset of CNO cycle in the shell. This boost leads to a high $^{14}$N production, which can be as much as $10^6$ times higher than the production of the non-rotating models. Generally, the rotating models produce much more metals than their non-rotating counterparts. The mass loss is very low, even for the models that reach the critical velocity during the main sequence. It may however have an impact on the chemical enrichment of the Universe, because some of the stars are supposed to collapse directly into black holes. They would contribute to the enrichment only through their winds. While in that case non-rotating stars would not contribute at all, rotating stars may leave an imprint in their surrounding. Due to the low mass loss and the weak coupling, the core retains a high angular momentum at the end of the evolution. The high rotation rate at death probably leads to a much stronger explosion than previously expected, changing the fate of the models. The inclusion of our yields in a chemical evolution model of the Galactic halo predicts log values of N/O, C/O and $^{12}$C/$^{13}$C ratios of -2.2, -0.95 and 50 respectively at log O/H $+12 = 4.2$.}
   {}


   \maketitle
%==============================================================================
\section{Introduction}

Population III (PopIII) stars occupy a key position in the evolution of the Universe. They are the first sources of light that have re-ionised the Universe. Also, they are the first producers of the heavy elements in the early Universe\footnote{According to the standard Big Bang Nucleosynthesis, only elements with nucleon number $A\leq 7$ are formed at a significant level during the first minutes of the Universe \citep[\emph{e.g.}][]{iocco07}.}.

Because of the peculiar conditions of the early Universe, especially the lack of coolants provided by the metals, the very first metal-free stars that formed are supposed to be very massive, with a typical mass $M\approx 10^2 - 10^3\ M_{\sun}$ \citep{abn02,bcl02}. \citet{nakum01} propose a bimodal IMF, with one peak at $M\approx 10^2\ M_{\sun}$ and the other at $M\approx 1\ M_{\sun}$. The bimodality is due to a threshold in the initial density of the cloud. Since this first suggestion that PopIII may not all be very massive objects, several studies show that the radiation field of a pre-existing star have a strong impact on subsequent star formation \citep{omyosh03, oshabwn05, greifbr06}. This would allow lower-mass metal-free stars to appear. Moreover \citet{silk06} show that even if the early Universe was free of magnetic fields, magnetic seeds can be generated in the disc surrounding an accreting stellar progenitor. These seeds can be amplified through a dynamo process and become strong enough to play an important role in the accretion rate. This again would reduce substantially the final mass range of the first stars. Though it has once been suggested that the most metal-poor stars known to this day may be low-mass PopIII that have undergone surface pollution \citep{shitsuyo03}, it is now widely accepted that metal-free stars were too massive to still be present nowadays ($M > 0.8 M_{\sun}$). The observation facilities are not yet able to reach far enough for a direct detection \citep[except maybe for the detection of pair-instability supernovae at intermediate redshifts, see][]{scan05}. The only way to constrain their physical properties is indirect observations of their chemical or radiative signature. Though it is not firmly established whether the most metal-deficient stars of the halo of our Galaxy are truly second generation stars \citep{salva07}, most of the authors consider they are born from a cloud enriched by only a few primordial progenitors \citep{christ02,umnom03}. Recently, accurate measurements of their surface abundances have started to allow some constraints to be put on the theoretical results of PopIII evolution \citep{cayrel04,barklem05,cohen07,aoki07}.

In view of their importance, PopIII stars have been the subject of many evolutionary calculations. Amongst the latest ones, most are presenting the evolution until AGB phase or central C ignition \citep{chieff01,mar01,mar03,siess02,gil07,lau08}. Others are focused on final stages \citep{HW02,UN05}. To our knowledge, there are two studies that include rotating models: the one by \citet{mar03}, and the one by \citet{hegerFS1}.  \citet{mar03} study very massive objects with masses between 120 and 1000 $M_{\sun}$. In their work, rotating stars are treated as rigid bodies and the effects of rotation intervenes mainly through a correction factor applied on the radiative mass loss (see Sect. \ref{sphymdot}). In these models, no rotational mixing is accounted for. \citet{hegerFS1} follow the evolution of differentially-rotating stellar models of $15-250\ M_{\sun}$ from the zero-age main sequence (ZAMS) to core collapse, but the detailed results of these models have not been published.

The present paper explores the effects of differential rotation on primordial massive stars. We think it is worth doing such a study for two main reasons. First, for metallicities between 0.004 and 0.040, the inclusion of the effects of differential rotation in stellar evolution codes has allowed to improve the agreement between the models and observations in many aspects: it has allowed to better reproduce the observed surface abundances \citep{heglang00,mm5}, the ratio of blue-to-red supergiants at low metallicity \citep{mm7}, the variation with the metallicity of the WR/O ratios and of the numbers of Type Ibc to Type II supernovae \citep{mm10,mm11}. It is interesting to see what a code that has been successfully confronted to observations can predict in a metallicity domain where no direct observations are available. Second, rotation has been shown to have important effects on low metallicity stars \citep{mem06}. During the Main Sequence (MS), the radiative winds are very low because of the lack of metals, so the star cannot get rid of its angular momentum. If it starts its evolution with a sufficiently high equatorial velocity, it may reach the critical limit. In that case, the outer layers are no more bound and some mass is lost, probably through a \emph{decretion} disc \citep{ow05} that may be quickly erased by the radiation. Further in the evolution, the rotational mixing brings freshly synthesised metals at the surface. The opacity of the outer layers are enhanced and the radiative mass loss is increased. Whereas the low metallicity stars were supposed to lose almost no mass at all, it has been shown that fast rotation can dramatically enhance the mass loss and thus modify drastically the evolution of these stars. How does a total absence of metals alter the above picture? It is interesting to extend the study to strictly $Z=0$ models.

The paper is organized as follow: in Sect. \ref{sphymod}, we describe the physical ingredients of the modelling; Sect. \ref{sevol} is dedicated to the presentation of the evolution of the models; in Sect. \ref{sfinstruc}, we present the pre-SN structure and the yields obtained; we study the effect of their inclusion in a galactic chemical evolution code in Sect.~\ref{sgalevol}, and then a discussion follows in Sect. \ref{sdiscu}.

Let us mention that electronic tables for the evolutionary sequences and the final structure files are available at \url{http://obswww.unige.ch/Recherche/evol/Rotating-primordial-stars}.
%==============================================================================
\section{Physical ingredients of the models \label{sphymod}}

The stellar models presented here have masses ranging from 9 to 200 $M_{\sun}$. The initial composition is $X=0.76$, $Y=0.24$\ and $Z=0$. The models with masses between 25 and 85 $M_{\sun}$ were computed until the end of the hydrostatic core Si burning. The 9 $M_{\sun}$\ model has developed a strongly degenerated core before carbon ignition and was thus stopped then. The 15 $M_{\sun}$\ model was stopped at the end of oxygen-burning also because of a too degenerate core at that time. Because of numerical difficulties, the 200 $M_{\sun}$ was stopped shortly after central He exhaustion.

The computation was carried out with the Geneva stellar evolution code. The version of the code is the one used in \citet{hmm12} that includes the necessary implementations to go on through advanced stages up to the pre-SN. It has been slightly modified to join the reaction networks of H and He burning during the MS, as required for zero-metallicity models.

The reaction rates are taken from NACRE database (\url{http://pntpm.ulb.ac.be/Nacre/barre_database.htm}). The opacities come from \citet{opa96}, complemented at low temperature by the molecular opacities of \citet{opamol94}. The treatment of convective instability is done according to Schwarzschild criterion. As in \citet{hir07}, convection is treated as a diffusive process from oxygen burning onwards. An overshooting parameter $\alpha_\mathrm{over}= 0.2\, H_\mathrm{P}$ is used for H- and He-burning cores only.
%-----------------------------------------------------------------------------------------------------------------------------------------
\subsection{Rotation \label{sphyrot}}
%:   tab tveloc
\begin{table}
\caption{Rotational properties of the models. The masses are in solar mass units, and the velocities in km~s$^{-1}$. The last column gives the central hydrogen mass fraction when the model arrives at the critical limit.}
\label{tveloc}
\centering
\begin{tabular}{r r r r r}
\hline\hline
    Mass & $\upsilon_\mathrm{ini}$ & $\upsilon/\upsilon_\mathrm{crit}$ & $\bar{\upsilon}\, {\rm (MS)}$  & $X_\mathrm{c, b}$ \\
\hline
        9 & 500 & 0.54 & 430 & -- \\
      15 & 800 & 0.72 & 626 & 0.018 \\
      25 & 800 & 0.63 & 657 & 0.253 \\
      40 & 800 & 0.56 & 700 & 0.326 \\
      60 & 800 & 0.52 & 740 & 0.360 \\
      85 & 800 & 0.48 & 754 & 0.399 \\
    200 & 800 & 0.40 & 733 & 0.428 \\
\hline
\end{tabular}
\end{table}

The effects of rotation are computed in the framework of the Roche model \citep[see][]{emmb08} and according to the shellular-rotation hypothesis \citep[see][]{z92}. The code couples the centrifugal force to the internal structure equations \citep{mmaraa}. This modifies the hydrostatic equilibrium and the surface conditions. The rotation-induced instabilities taken into account are the meridional circulation, the secular and dynamical shears.
\begin{itemize}
\item The meridional circulation arises from the impossibility of a rotating star to be in hydrostatic and radiative equilibrium at the same time (von Zeipel theorem). The vertical component of the meridional circulation velocity is $u(r,\theta)=U(r)\,P_2(\cos \theta)$ with $P_2(x)$ the second Legendre polynomial. The formulation of the radial amplitude $U(r)$ has been determined by \citet{z92} and \citet{mz98} as:
\begin{eqnarray}
U(r)&=&\frac{P}{\rho g C_P T \left[ \nabla_{\rm ad} - \nabla_{\rm rad} + (\varphi/\delta) \nabla_{\mu} \right]} \nonumber \\
 & &\hspace{1cm}\cdot \left( \frac{L}{M_\star} 
\left[ E_\Omega + E_\mu \right] + \frac{C_P}{\delta} \frac{\partial \Theta}{\partial t} \right) \nonumber
\end{eqnarray}
where $C_P$ is the specific heat, $\nabla_{\rm ad}=\frac{P \delta}{\rho T C_P}$ the adiabatic gradient, $M_\star=M \left(1 - \frac{\Omega^2}{2\pi g \rho_{\rm m}} \right)$, and $\Theta=\tilde{\rho}/\bar{\rho}$ is the ratio of the variation of the density to the average density on an equipotential. $\varphi$ and $\delta$ arise from the generic equation of state: $\frac{d \rho}{\rho} = \alpha \frac{d P}{P} + \varphi \frac{d \mu}{\mu} - \delta \frac{d T}{T}$. $E_\Omega$ and $E_\mu$ are terms which depend on the $\Omega$- and $\mu$-distributions respectively \citep[see][for details on these expressions]{mz98}. $E_\Omega$ is the dominant term in the big parenthesis, and it can be reasonably approximated by: $E_\Omega \simeq \frac{8}{3} \left[ 1- \frac{\Omega^2}{2\pi g \bar{\rho}} \right] \left( \frac{\Omega^2 r^3}{GM} \right)$. The term in brackets is known as the \emph{Gratton-\"Opik term}. It may become largely negative in the outermost layers of the stars, where $\bar{\rho}$ becomes very small, which leads to $U(r) < 0$. In this case, the circulation goes down along the polar axis from the surface inwards, and then rises in the equatorial plane outwards to the surface, bringing angular momentum to the surface.
\item Differential rotation induces shear turbulence at the interface of layers having different rotational velocities. A layer remains stable if the excess kinetic energy due to the differential rotation does not overcome the buoyancy force. It is known as the \emph{Richardson criterion} and is expressed as $Ri = \frac{N^2}{\left(\partial u/\partial z \right)^2} > Ri_{\rm c}$, where $u$ is the horizontal velocity, $z$ the vertical coordinate, and $N^2=\frac{g\delta}{H_P}\left( \nabla_{\rm ad} - \nabla + (\varphi/\delta)\nabla_{\mu} \right)$ is the Brunt-V\"ais\"al\"a frequency. $Ri_{\rm c}=1/4$ is the critical value for stability. The coefficient of diffusion by dynamical shear implemented in the code is described in \citet{hmm12}:
$$
D_{\rm dyn} = \frac{1}{3}\, r\, \Delta \Omega\, \Delta r
$$
where $r$ is the mean radius of the zone constituted of consecutive shells with $Ri < Ri_{\rm c}=1/4$, $\Delta \Omega$ is the variation of $\Omega$ over this zone and $\Delta r$ is the extend of the zone. The timescale for the dynamical shear is just as it sounds: the dynamical timescale $\tau_{\rm dyn}=\sqrt{r^3/(Gm_r)}$. The prescription for $D_{\rm dyn}$ is valid when the thermal adjustment timescale $\tau_{\rm th}=\frac{d^2}{12\,K}$ (for a sphere with diameter $d$ and a thermal diffusivity $K$) is longer than the dynamical timescale.
\item Secular shear arises when the thermal dissipation reduces the buoyancy force. The coefficient of diffusion by secular shear turbulence is determined by \citet{Maed97} as:
\begin{eqnarray}
D_{\rm shear}& = & \frac{4 \left(K+D_h\right)}{\left[\frac{\varphi}{\delta}\nabla_{\mu}\left(1+\frac{K}{D_h}\right) + \left( \nabla_{\rm ad} - \nabla_{\rm rad} \right)\right]} \nonumber \\
 & & \hspace{1cm}  \cdot \frac{H_P}{g\delta} \left[ \frac{ \alpha}{4} \left( 0.8836\ \Omega\ \frac{d \ln \Omega}{d \ln r} \right)^2 - \left(\nabla' - \nabla \right) \right] \nonumber
\end{eqnarray}
with $K=\frac{4acT^3}{3\kappa \rho^2 C_P}$ the thermal diffusivity, $\alpha$ the fraction of the excess energy in the shear that will contribute to mixing, and $(\nabla' - \nabla)$ the difference between the internal non-adiabatic gradient and the local gradient. The latter term is extremely small and is neglected in the calculation. $D_{h}$ is the coefficient for the horizontal turbulence. It is determined as in \citet{z92}:$D_h = \frac{1}{c_h}\ r\ \left| 2\,V(r) - \alpha\,U(r) \right|$, where $c_h$ is a constant of order 1, $V(r)$ is the horizontal component of the meridional circulation velocity, $U(r)$ its vertical component, and in this expression $\alpha = \frac{1}{2} \frac{d \ln (r^2 \bar{\Omega})}{d \ln r}$.
 \end{itemize}
In the zones where $Ri < Ri_{\rm c}$, the shear mixing coefficient is eventually set to:
$$
D_{\rm shear} = D_{\rm dyn} + D_{\rm shear}
$$

The transport of angular momentum inside a star is implemented following the prescription of \citet{z92}, complemented by the works of \citet{tz97} and \citet{mz98}. In the radial direction, it obeys the following equation:
$$
\rho \frac{d}{dt} \left( r^2 \bar{\Omega} \right)_{M_r} =
   \frac{1}{5r^2} \frac{\partial}{\partial r} \left( \rho r^4 \bar{\Omega} U(r) \right)
   + \frac{1}{r^2} \frac{\partial}{\partial r} \left( \rho D r^4 \frac{\partial \bar{\Omega}}{\partial r} \right)
$$
The first term in the right hand side of this equation is the divergence of the \emph{advected} flux of angular momentum, while the second term is the divergence of the \emph{diffused} flux. The coefficient $D$ is the total diffusion coefficient, taking into account the various instabilities which transport angular momentum (convection, shears). After the end of the MS, the timescale of the evolution shortens, so only the diffusive term is kept in the formulation of the transport of the angular momentum.

The transport of chemical elements by the meridional circulation has been shown to be satisfactorily approximated by a diffusion process \citep[see][]{chabz92} throughout the whole evolution.

Since very low- or zero-metallicity stars are more compact than their metal-rich counterparts, a similar ratio of the initial velocity to the critical velocity\footnote{The critical velocity $\upsilon_\mathrm{crit}$\ is reached when the gravitational acceleration is exactly counterbalanced by the centrifugal force. In the framework of the Roche model, one has: $\upsilon_\mathrm{crit}= \sqrt{\frac{2}{3}\frac{GM}{R_\mathrm{pb}}}$, where ${R_\mathrm{pb}}$\ is the polar radius at break-up limit.} $\upsilon/\upsilon_\mathrm{crit}$ leads to a higher equatorial velocity. In order to explore the effects of rotation on metal-free models, we chose a high initial velocity, 800 km~s$^{-1}$. Note that for the 60 $M_{\sun}$\ model, this initial velocity corresponds to a total angular momentum content that is similar to a solar metallicity model of the same mass rotating at 300 km~s$^{-1}$. The 9 $M_{\sun}$\ model was computed with a lower initial velocity (500 km~s$^{-1}$) because 800 km~s$^{-1}$ is overcritical on the ZAMS for this model. Table \ref{tveloc} presents the initial equatorial velocities of the models, the corresponding $\upsilon/\upsilon_\mathrm{crit}$ ratio and the mean velocity during MS.

In order to be able to compare our results with other works, we computed, for each mass, a corresponding model without rotation.
%-----------------------------------------------------------------------------------------------------------------------------------------
\subsection{Mass loss \label{sphymdot}}
Zero-metallicity stars are supposed to evolve at constant mass due to a scaling of the wind with metallicity of the form $\dot{M} \propto (Z/Z_{\sun})^{0.5}$. However, according to \citet{kudr02}, H and He lines are able to remove some mass if the luminosity is sufficiently high. We thus adopted the \citet{kudr02} prescription for the models presenting a luminosity $\log (L/L_{\sun}) > 6$. The 85 and 200 $M_{\sun}$\ models were computed with it from the start, while we turned it on during the MS evolution of the 60 $M_{\sun}$. Since Kudritzki's prescription does not correspond to the strictly $Z=0$\ case, we used the same adaptations as in \citet{mar03}: 1) the surface metallicity is set $Z=\max(Z_\mathrm{min}, Z_\mathrm{eff})$, where $Z_\mathrm{min}=10^{-4}\,Z_{\sun}$ is the lowest value in the validity range of Kudritzki's prescription, and $Z_\mathrm{eff}=1-X-Y$ is the effective surface metallicity of the model (allowing for self-enrichment); and 2) the effective temperature is set $T_\mathrm{eff}=\min(T_\mathrm{eff}, 60'000\,K)$, \emph{i.e.} we did not extrapolate Kudritzki's formula beyond its validity domain but constrained $T_\mathrm{eff}\leq 60'000\,K$ for the calculation of the mass loss. The models with a mass lower than 60 $M_{\sun}$ never reach $\log (L/L_{\sun}) > 6$, so they were computed without radiative mass loss.
%
%:   fig fdhr
   \begin{figure*}
   \centering
 \resizebox{\hsize}{!}{\includegraphics{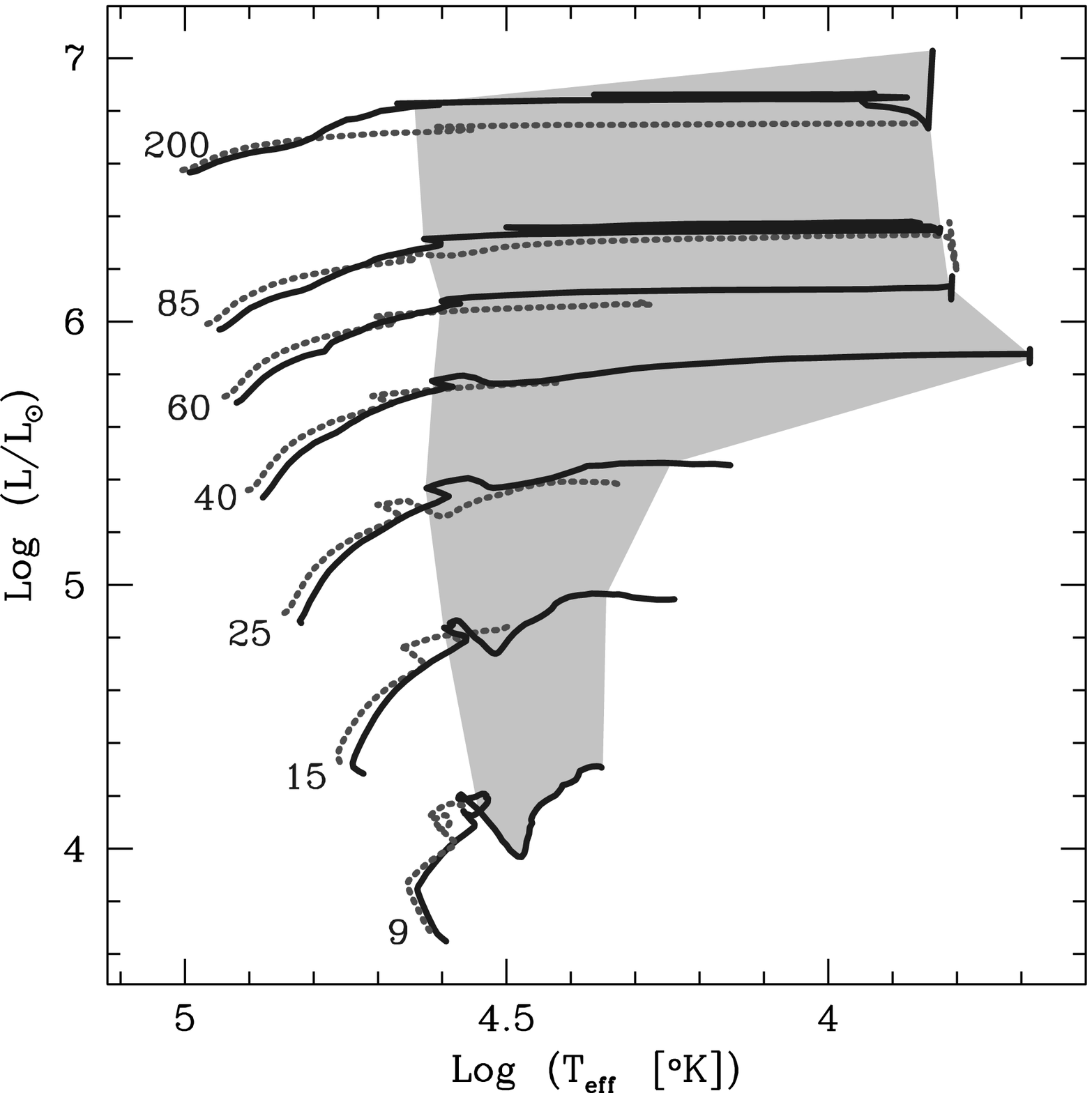}\hspace{.2cm} \includegraphics{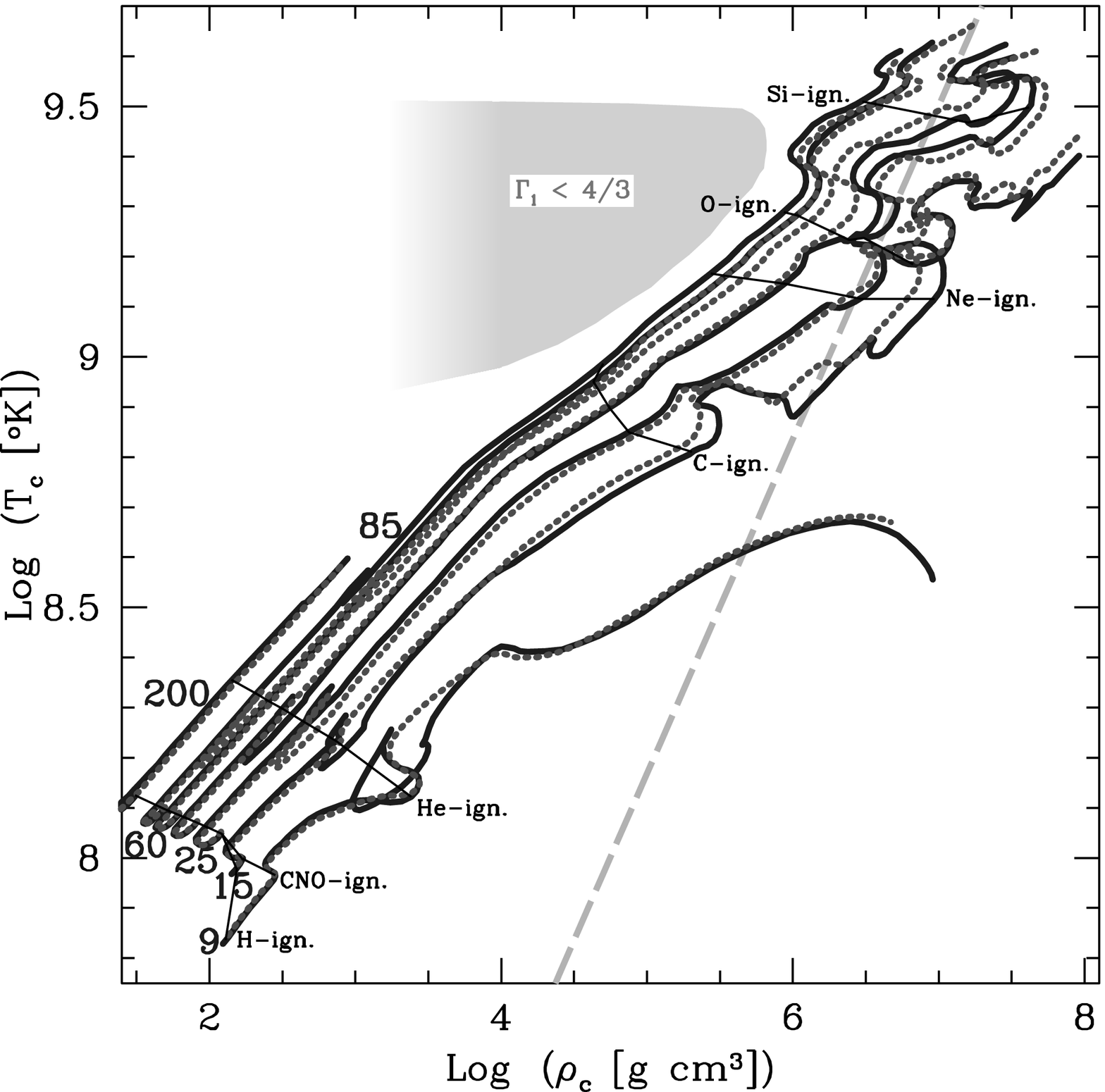}}
      \caption{Evolution of $Z=0$ models (with rotation: solid lines; without rotation: dotted lines). {\it Left:} Hertzsprung Russell diagram. The grey area shows the zone of the diagram where He burns in the core of the rotating models. {\it Right:} central temperature vs central density diagram. The ignition of the different burning stages are given for the rotating models. The grey area indicates the zone of the diagram where the pair-creation instability occurs. The dashed grey line shows the locus of the points where the perfect gas pressure is equal to the completely-degenerate non-relativistic pressure.}
         \label{fdhr}
   \end{figure*}

Rotation, by lowering the effective gravity at the surface, helps the radiation removing mass from the star. As shown in \citet{mm6}, the line-driven mass loss is enhanced by a factor which depends on the angular velocity such that the mass loss in the case of rotation becomes: $$\dot{M}(\Omega)= F_{\Omega}\cdot \dot{M}(\Omega=0) \hspace{.3cm}
\mathrm{where}\hspace{.2cm} F_{\Omega}=\frac{(1-\Gamma)^{\frac{1}{\alpha}-1}}{\left[ 1-\frac{\Omega^2}{2\pi G \rho_\mathrm{m}} - \Gamma \right]^{\frac{1}{\alpha}-1}}$$
We see that this expression depends also on the Eddington factor $\Gamma=L/L_\mathrm{Edd}=\kappa L / (4\pi cGM)$ where $\kappa$ is the electron-scattering opacity.

Moreover, there is another situation in which the models can lose mass: it is when the external layers of the star reach the critical velocity. The over-critical layers are no more bound to the star, though it is not yet clear whether they are expelled in the interstellar medium or remain in the surrounding, maybe in the form of a disc. However, the removal of these layers brings the surface back to sub-critical velocity until the evolution of the star accelerates it again. In the present calculation, as in \citet{mem06}, once the model reaches the break-up limit, we adapted the mass loss rate to keep the surface just under the critical velocity. This treatment was applied to all the models except the 9 $M_{\sun}$, which never reached the critical limit.
%==============================================================================
\section{Evolution \label{sevol}}
%-----------------------------------------------------------------------------------------------------------------------------------------
\subsection{Hertzsprung Russell diagram and central conditions \label{sdhr}}
%:   fig fflash
   \begin{figure*}
   \centering
 \resizebox{\hsize}{!}{\includegraphics{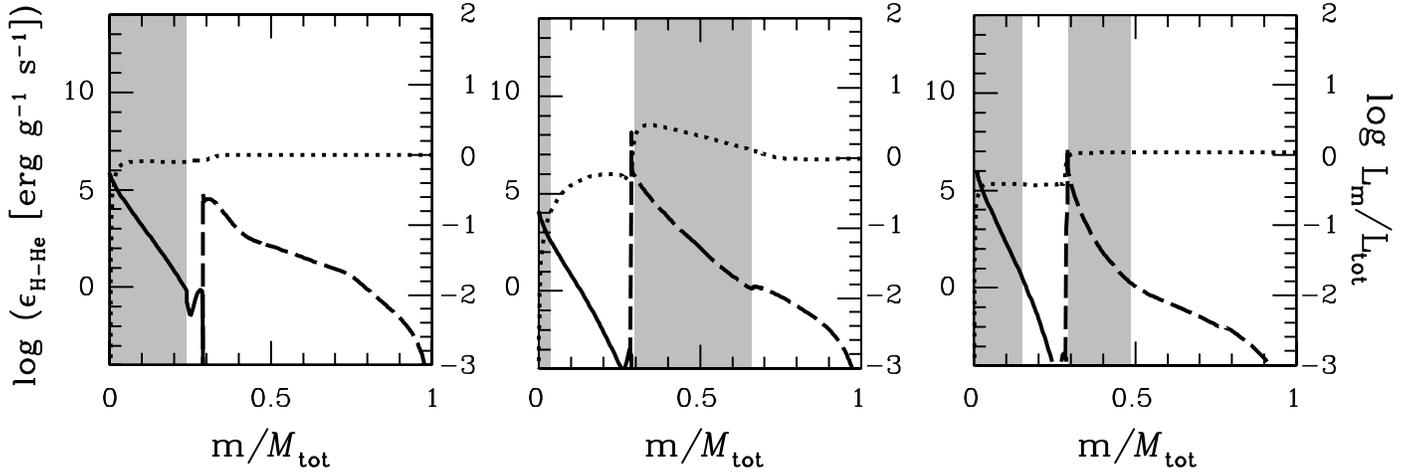}}
      \caption{Internal structure of the 15 $M_{\sun}$ rotating model at different stages of the core He burning. The solid line is the He-burning energy production and the dashed line the H-burning energy production. The dotted line is the fraction of the total luminosity (scale on the right). The grey areas show the convective zones. \emph{Left:} before the CNO shell boost (age: 13.739 Myr; $Y_\mathrm{c}= 0.4632$) - \emph{centre:} at the moment of the boost (age: 13.744 Myr; $Y_\mathrm{c}= 0.4919$) - \emph{right:} after the boost (age: 13.757 Myr; $Y_\mathrm{c}= 0.4840$).}
         \label{fflash}
   \end{figure*}

In the left panel of Fig.~\ref{fdhr}, we present the Hertzsprung Russell diagram (HRD) of our rotating and non-rotating models, and in the right panel, their tracks in the $\log T_\mathrm{c} - \log \rho_\mathrm{c}$ diagram. We notice the expected characteristics of rotating models: the ZAMS is shifted towards lower effective temperature and luminosity with respect to the non-rotating case\footnote{We recall that this shift is due to the sustaining effect of the rotation: the gravity is counter-balanced both by the gas pressure and the centrifugal force in such a way that the star behaves like a lower mass one.}. Then, when the evolution proceeds, the tracks become more luminous, and the main-sequence turn-off is shifted to cooler temperature: the core of the rotating models is refuelled by fresh H brought by the mixing. It thus grows, leading to an enhancement of the luminosity.

The 9 and 15 $M_{\sun}$ models are powered only by $pp$-chains when they arrive on the ZAMS and thus they continue their initial contraction. It is only when they have produced a sufficient amount of carbon through $3\alpha$ reaction (about $10^{-12}$ in mass fraction) that the CNO cycle can start, drawing a hook in the $\log T_\mathrm{c} - \log \rho_\mathrm{c}$ diagram. The onset of the CNO cycle in these models can also be seen in the HRD:  their tracks evolve towards the blue side of the diagram, until the energy provided by the CNO cycle stops the contraction and bends the tracks back in the usual MS feature. In the rotating 9 $M_{\sun}$, this happens at an age of 12.2 Myr (when the central H mass fraction is $X_\mathrm{c}=0.439$) while in the non-rotating one it happens a little earlier, at an age of 10.9 Myr (but at a similar burning stage: $X_\mathrm{c}=0.439$). In the case of the non-rotating 15 $M_{\sun}$ model, it happens after merely 1.5 Myr ($X_\mathrm{c}=0.695$), while it takes 2 Myr ($X_\mathrm{c}=0.677$) in the case of the rotating one. Let us mention that \citet{mar01} find that the mass limit for the CNO ignition to occur already on the ZAMS is 20 $M_{\sun}$. Our results are consistent with that limit.

After central H exhaustion, the core He-burning phase (CHeB) starts right away: the core was already hot enough to burn a little He during the MS so it does not need to contract much further. The transition between core H burning and He burning is smooth because the model is continuously sustained by core nuclear burning. This prevents the models to start a redward evolution, so they remain in the blue part of the HRD at the beginning of CHeB. The H-burning shell is powered only by pp-chains and remains radiative. The main part of the luminosity is provided by the core (see in Fig.~\ref{fflash} \emph{left} the case of the rotating 15 $M_{\sun}$). Something particular happens to the rotating models during the CHeB phase: because of rotational mixing, some carbon produced in the core is diffused towards the H-burning shell, allowing a sudden ignition of the CNO cycle in the shell. This boost of the shell leads to a retraction of the convective core and a decrease of the luminosity. At the same time, it transforms the quiet radiative H-burning shell into an active convective one (see Fig.~\ref{fflash} \emph{centre}). Later, the model takes the structure of a higher metallicity one: convective He-burning core, convective H-burning shell and expanding envelope (see Fig.~\ref{fflash} \emph{right}). This boost appears in Fig.~\ref{fdhr} (\textit{left}) as a deep V-shaped feature in the rotating tracks of the 9 and 15 $M_{\sun}$, and as a slight depression in the 25 and 40 $M_{\sun}$ tracks: the luminosity decreases at the same time as the $T_\mathrm{eff}$ cools down. The only non-rotating models that present the same feature are the 25 and 85 $M_{\sun}$. In this case, the structure is such that the growing convective core reaches the bottom of the quiet radiative H-burning shell, triggering the same event than in the rotating models.  Such a CNO boost leaves an imprint in the $\log T_\mathrm{c} - \log \rho_\mathrm{c}$ diagram: a second hook can be seen in the tracks, which takes the shape of a complex loop in the case of the 9 $M_{\sun}$.

The crossing of the HRD is slow, contrarily to higher metallicity models. As we mentioned above, the nuclear burning is continuous inside the core, therefore the $Z=0$ models do not go through a phase of rapid core contraction and its mirror effect of fast envelope expansion. On the HRD, we note that the crossing of the diagram takes place during the whole CHeB (grey area).

In the non-rotating set, all the models with mass between 9 and 60 $M_{\sun}$ end their life as blue supergiants (BSG), while the 85 and 200  $M_{\sun}$ become red supergiants (RSG). This compares well with the tracks obtained by \citet{mar01}. As in their study, all the models with masses between 9 and 60 start and finish core He burning in the blue part of the HRD. Only the highest masses models ($M\geq 70\ M_{\sun}$) end their central He combustion on their Hayashi line.

Generally, the rotating tracks end their evolution in a cooler part of the HRD than their non-rotating counterparts. The only exception is the 85 $M_{\sun}$, which ends its life as a BSG because of a blue loop after CHeB. The rotating 15 and 25 $M_{\sun}$ end their evolution in the blue part of the HRD, without any redward incursion. The rest of the rotating models become RSG. Let us note that none of our models meet the conditions to be Wolf-Rayet stars. The 85 $M_{\sun}$ ends its life with a $\log T_\mathrm{eff}$ higher than 4.3, but its surface abundance of hydrogen remains too high.
%-----------------------------------------------------------------------------------------------------------------------------------------
\subsection{Lifetimes \label{stau}}
%:   tab ttaucores
\begin{table}
\caption{Final masses, total lifetime, MS and core He-burning durations, masses of the He- and CO-core for our models at the pre-SN stage. Masses are in $M_{\sun}$, velocities in km s$^{-1}$ and times in Myr.}
\label{ttaucores}
\centering
\begin{tabular}{r r r r r r r r}
\hline\hline
    Mass & $\upsilon_\mathrm{ini}$ & $M_\mathrm{fin}$ & $\tau_\mathrm{life}$ & $\tau_\mathrm{H}$ & $\tau_\mathrm{He}$ & $M_\mathrm{He}$ & $M_\mathrm{CO}$\\
\hline
    9 &     0 &  ~9.00 & 22.0 & 19.50 & 1.99 & 1.73 & 0.46 \\
    9 & 500 &  ~9.00 & 27.1 & 23.90 & 2.58 & 2.09 & 0.63 \\
    15 &     0 &  15.00 & 11.6 & 10.60 & 0.83 & 3.62 & 2.92 \\
    15 & 800 &  14.96 & 14.4 & 13.20 & 0.98 & 4.31 & 2.71 \\
    25 &     0 &  25.00 & 7.29 & 6.64 & 0.56 & 7.51 & 5.39 \\
    25 & 800 &  24.75 & 8.47 & 7.88 & 0.50 & 9.14 & 6.30 \\
    40 &     0 &  40.00 & 5.03 & 4.60 & 0.36 & 15.95 & 14.52 \\
    40 & 800 &  37.99 & 5.88 & 5.41 & 0.41 & 17.22 & 13.53 \\
    60 &     0 &  59.99 & 3.96 & 3.60 & 0.30 & 26.52 & 25.11 \\
    60 & 800 &  57.59 & 4.57 & 4.20 & 0.31 & 30.86 & 30.58 \\
    85 &      0 &  84.93 & 3.42 & 3.05 & 0.31 & 34.77 & 34.50 \\
    85 & 800 &  74.57 & 3.84 & 3.52 & 0.28 & 44.38 & 43.92 \\
    200 &     0 &199.95 & 2.53 & 2.27 & 0.23 & 101.63 & 95.69 \\
    200 & 800 &183.56 & 2.83 & 2.56 & 0.24 & 109.65 & 95.97 \\
\hline
\end{tabular}
\end{table}

Table~\ref{ttaucores} shows the lifetimes for the MS and CHeB phases. During MS, rotation induces an increase of the lifetime from 12.8\% for the 200 $M_{\sun}$ up to 24.5\% for the 15 $M_{\sun}$. The models with higher initial $\upsilon/\upsilon_\mathrm{crit}$ show a more important increase. For the models that reach the critical limit, we find an almost linear relation between the relative enhancement of the MS lifetime $\Delta \tau_\mathrm{H}$ and the initial ratio $\upsilon/\upsilon_\mathrm{crit}$ of the form
%:   fig furvevolz
   \begin{figure*}
   \centering
    \resizebox{\hsize}{!}{\includegraphics{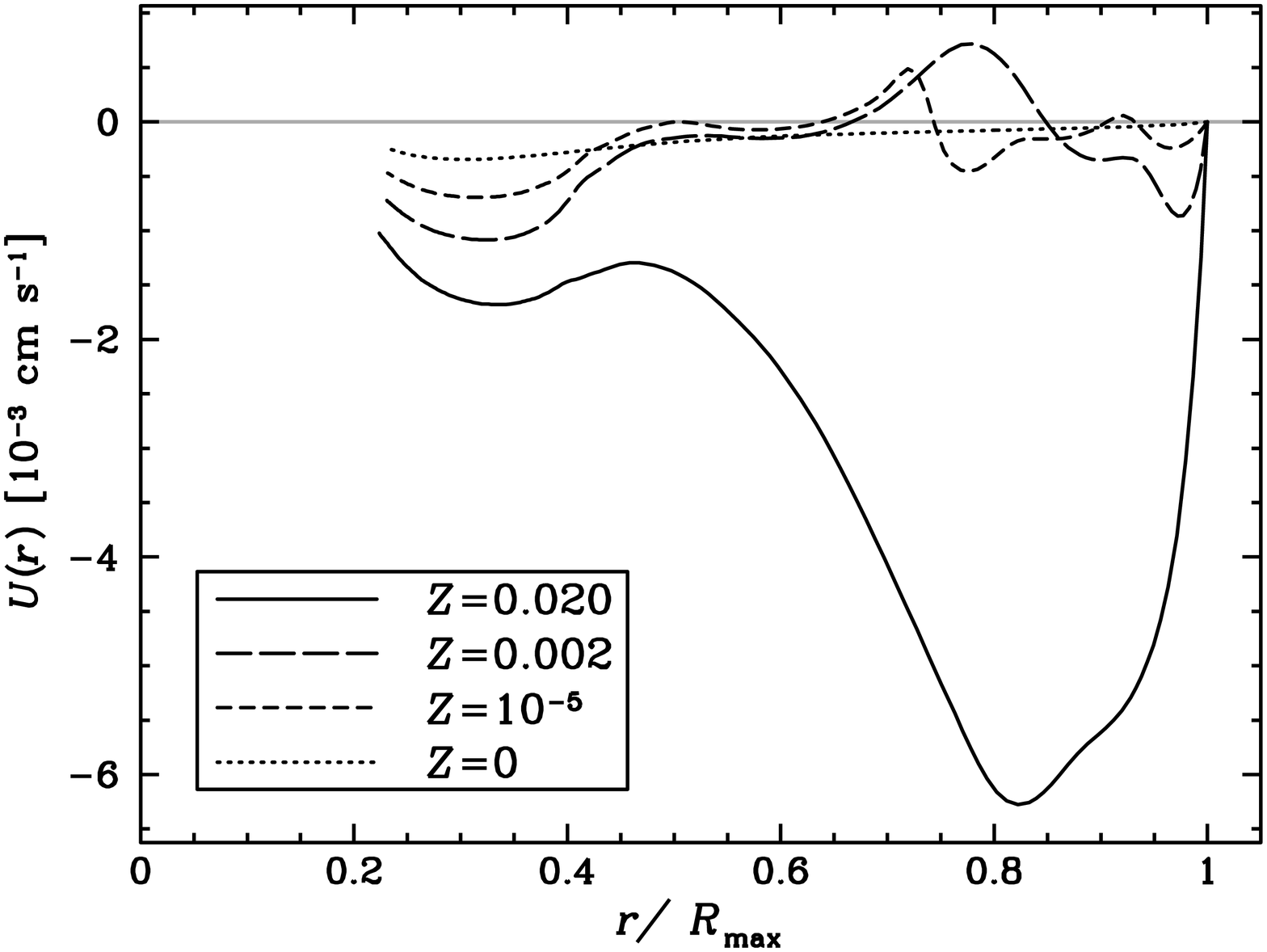}\hspace{.2cm} \includegraphics{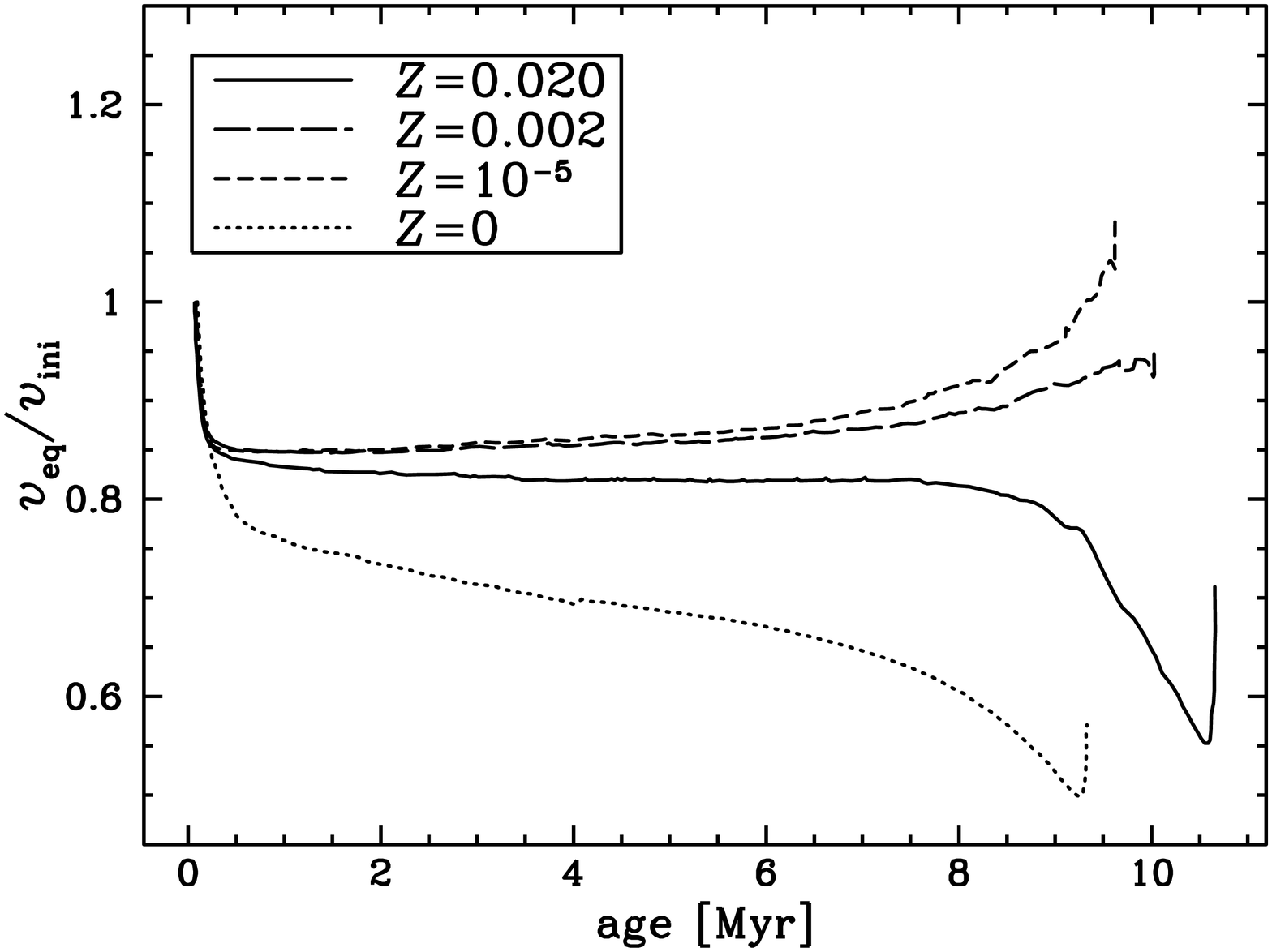}}
      \caption{Models of 20 $M_{\sun}$ at various metallicities, with $\Omega_\mathrm{ini}/\Omega_\mathrm{crit}= 0.5$. {\it Left:} internal profile of $U(r)$, where $u(r,\theta)$ the vertical component of the velocity of the meridional circulation is $u(r,\theta)= U(r)\ P_2(\cos \theta)$ with $P_2$ the second Legendre polynomial. The radius is normalized to the outer one. All the models are at the same evolutionary stage, when the central H mass fraction is about 0.40. {\it Right:} evolution of the equatorial velocity, normalized to the initial velocity.}
         \label{furvevolz}
   \end{figure*}
 $$
 \Delta \tau_\mathrm{H} = \frac{\tau_\mathrm{H}(\upsilon/\upsilon_\mathrm{crit}) - \tau_\mathrm{H}(0)}
                                                      {\tau_\mathrm{H}(0)} = 0.362\ \upsilon/\upsilon_\mathrm{crit} 
 $$
(when $\tau_\mathrm{H}$ is expressed in Myr) which reproduces the values of $\Delta \tau_\mathrm{H}$ obtained in our models with an accuracy better than 7\%.
 
How does our work compare with the other studies? For our non-rotating 9 $M_{\sun}$, the MS lifetime is shorter by 12\% than that found by \citet{mar01}, which is consistent with our lower overshoot parameter: the convective core of our model is about 10\% smaller than theirs. Let us mention also that they have a higher initial hydrogen abundance (X=0.77). \citet{gil07} provide two models of 9 $M_{\sun}$, one computed without overshoot and the other with the overshoot theory presented by \citet{et04}. Their calculation of the convective boundaries could be approximated by $\alpha_\mathrm{over}= 0.3\, H_\mathrm{P}$ (P. Gil-Pons, priv. comm.). The lifetime of our model is about 21\% shorter than the model they calculated with this higher overshoot. But it is also 11\% shorter than their model calculated without any overshoot. In a previous paper \citep{gil05}, they show the MS evolution of the convective core and of the H and He abundances for this model. We notice that the transition from $pp$-chain burning to CNO cycle occurs earlier in their model (4.4 Myr) than in ours (10 Myr), and in a much more abrupt way (see their Fig.~4). The convective core expands suddenly, the H abundance rises in the core and thus allows a longer lifetime. The difference in the time at which the CNO cycle ignites arises most probably from differences in the reaction rates and in the equation of state used.

In the case of our 15 $M_{\sun}$ non-rotating model, we find again a difference of about 10\% with the work of \citet{mar01}, consistent with the difference found in the 9 $M_{\sun}$ models. Our lifetime is also 12\% shorter than that calculated by \citet{siess02} for their (non-rotating) 15 $M_{\sun}$, although the overshooting parameter they use is the same as ours. Since their paper is focused on low- and intermediate-mass stars, we cannot find in it any explanation for this difference.

In order to compare our results with the one of \citet{mar03} at higher masses, we computed the evolution of a non-rotating 120 $M_{\sun}$ up to the end of the MS. Once more the MS lifetime
of our model is found to be 10\% shorter than the one computed by the latter authors.
%-----------------------------------------------------------------------------------------------------------------------------------------
\subsection{Evolution of the equatorial velocity \label{sveq}}

Our models are started on the ZAMS with a flat rotation profile. As described in \cite{mm5}, the meridional circulation establishes then very quickly (1-2\% of the MS time) an equilibrium profile of $\Omega(r)$ inside the star with the core spinning faster than the surface.

After this first adjustment phase, the evolution of the surface (equatorial) velocity results of a delicate interplay between the mass loss (which removes angular momentum at the surface) and the meridional circulation (which brings angular momentum from the core to the surface). How does this interplay behave at $Z=0$, and what are the differences with the situation at higher metallicities? To find an answer, we compared four models of 20 $M_{\sun}$ with $\Omega_\mathrm{ini}/\Omega_\mathrm{crit}= 0.50$ at various metallicities, models which were computed with the same code as our set of $Z=0$ models\footnote{These models are taken from \citet{emmb08}}. In the left panel of Fig.~\ref{furvevolz}, we show the internal profile of the radial component of the meridional circulation. Let us focus on the outer cell, which transports the angular momentum outward, when $U(r)$ is negative. The amplitude of the meridional circulation is a factor 6 higher in the standard metallicity model ($Z=0.020$) than in the $Z=0.002$ one. This factor amounts to 25 when we compare with the $Z=10^{-5}$ model, and reaches 100 with the $Z=0$ one. This illustrates the effect of the Gratton-\"Opik term (see Sect.~\ref{sphyrot}) in the expression of the meridional circulation velocity $U(r)$: when the metallicity decreases, stars are more compact, so the density increases, and thus $U(r)$ decreases. In the right panel of Fig.~\ref{furvevolz}, we see the resulting evolution of the equatorial velocity. At standard metallicity, although the amplitude of the meridional circulation is large, the loss of angular momentum through the radiative winds has the strongest effect, and the equatorial velocity slows down. At low or very low $Z$, the meridional circulation is weak, but the mass loss is so diminished that the models are spinning up. In the case of $Z=0$ strictly, there is no mass loss to remove mass and angular momentum, but the meridional circulation is so weak that the evolution of $\Omega(r)$ is very close to local angular momentum conservation, $\Omega r^2 = \mathrm{constant}$: because of the natural inflation of the external radius, the surface of the model has to slow down.
%:   tab tmlost
\begin{table*}
\caption{Total mass lost and surface abundances of C, N and O  at the end of MS phase and at the end of the evolution for the models that undergo mass loss and surface enrichment. Masses are in $M_{\sun}$, velocities in km s$^{-1}$ and abundances in mass fraction. The C abundance is the sum of $^{12}$C and $^{13}$C, the N abundance is the abundance in $^{14}$N and the O abundance is the sum of $^{16}$O, $^{17}$O and $^{18}$O.}             
\label{tmlost} 
\centering
\begin{tabular}{r r r c c c c r c c c c}
\hline\hline
 $M_\mathrm{ini}$ & $\upsilon_\mathrm{ini}$ & \multicolumn{5}{|c|}{end of MS} & \multicolumn{5}{|c|}{end of evolution} \\
  &  & $M_\mathrm{lost}$ & He & C & N & O & $M_\mathrm{lost}$ & He & C & N & O \\
\hline 
   9 & 500 & --\ \ \ \ \  & 0.2426 & 4.60e-23 & 5.91e-21 & 2.35e-22 & --\ \ \ \ \  & 0.4279 & 5.50e-17 & 5.61e-15 & 1.85e-16 \\
  15 & 800 & 0.0306 & 0.2503 & 1.74e-14 & 1.81e-12 & 5.90e-14 & 0.0418 & 0.2646 & 6.36e-09 & 1.83e-07 & 3.38e-09 \\
  25 & 800 & 0.2252 & 0.2517 & 4.34e-13 & 4.02e-11 & 1.20e-12 & 0.2513 & 0.2615 & 3.53e-10 & 9.37e-09 & 1.86e-10 \\
  40 & 800 & 1.0070 & 0.2654 & 2.24e-12 & 1.90e-10 & 5.18e-12 & 2.0074 & 0.3045 & 5.17e-07 & 1.48e-05 & 6.16e-07 \\
  60 &     0 & --\ \ \ \ \  & 0.2400 & 0.00e+00 & 0.00e+00 & 0.00e+00 & 0.0001 & 0.2400 & 0.00e+00 & 0.00e+00 & 0.00e+00 \\
  60 & 800 & 2.3200 & 0.2906 & 6.84e-12 & 5.44e-10 & 1.39e-11 & 2.4061 & 0.5328 & 1.23e-05 & 3.54e-04 & 3.34e-05 \\
  85 &     0 & 0.0046 & 0.2400 & 0.00e+00 & 0.00e+00 & 0.00e+00 & 0.0681 & 0.4032 & 6.40e-04 & 3.88e-03 & 1.33e-04 \\
  85 & 800 & 4.4363 & 0.3362 & 1.74e-11 & 1.31e-09 & 3.19e-11 & 10.4239 & 0.6141 & 7.29e-06 & 1.81e-04 & 4.14e-05 \\
200 &     0 & 0.0350 & 0.2400 & 0.00e+00 & 0.00e+00 & 0.00e+00 & 0.0509 & 0.2423 & 7.93e-13 & 6.12e-11 & 1.59e-12 \\
200 & 800 & 16.4426 & 0.5183 & 8.82e-11 & 5.94e-09 & 1.23e-10 & 20.7609 & 0.6685 & 7.53e-05 & 2.09e-04 & 1.02e-04 \\
\hline
\end{tabular}
\end{table*}

The behaviour described above is valid for all the models with $M_\mathrm{ini} < 40\ M_{\sun}$: they are slowing down during the MS evolution, except at the very end of it when they contract briefly before central H exhaustion. Above 40 $M_{\sun}$, the models are spinning up on the MS. This is not surprising, since the meridional circulation is higher in more massive stars (which have a less dense envelope), allowing the transport of angular momentum that is lacking in lower mass models.

%:   fig fooc
   \begin{figure}
   \centering
    \resizebox{\hsize}{!}{\includegraphics[width=\textwidth]{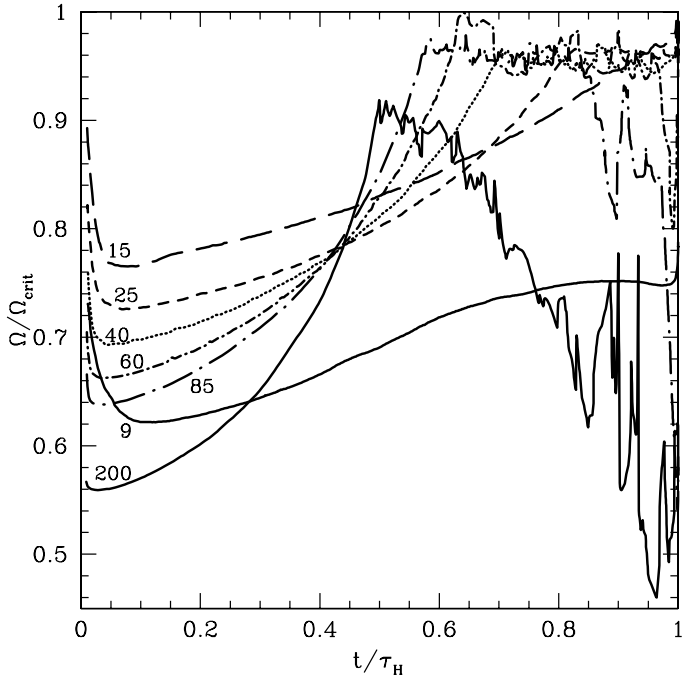}}
      \caption{Evolution of the $\Omega/\Omega_\mathrm{crit}$ ratio during the MS. All the models start the MS with $\upsilon_\mathrm{eq}=800$ km s$^{-1}$, except the 9  $M_{\sun}$ which starts the MS at 500 km s$^{-1}$.}
         \label{fooc}
   \end{figure}

An important remark has to be made here: during the evolution, because of the inflation of the radius, the value of the critical velocity gets lower. So even a slowing down model may reach the critical velocity at a time of its evolution. This is indeed what happens with our models. In Fig.~\ref{fooc} we present the evolution of the $\Omega/\Omega_\mathrm{crit}$ ratio during the MS. As mentioned in Sect.~\ref{sphyrot}, the 9  $M_{\sun}$ model is the only one that does not reach the critical limit. For the other models, the higher the mass the sooner the critical limit is attained. The last column of Table~\ref{tveloc} gives the central mass fraction of hydrogen when each model reaches the critical velocity.

Once at critical limit, all the models remain at the limit. In Fig.~\ref{fooc}, the 85 and especially the 200 $M_{\sun}$ seem to depart from $\Omega/\Omega_\mathrm{crit}=1$, but this is because the limit shown here is only the $\Omega$-limit, where the centrifugal force alone is taken into account to counterbalance the gravity. In the two above models however, the radiative acceleration is strong and they reach the so-called $\Omega\Gamma$-limit, that is the second root of the equation giving the critical velocity: $\vec{g_\mathrm{eff}}\,[1-\Gamma]=\vec{0}$ \citep[see][for the detailed derivation]{mm6}. The true critical velocity is lowered by the radiative acceleration, and though the plotted $\Omega/\Omega_\mathrm{crit}$ ratio becomes lower than 1, these models are actually at the critical limit and remain at this limit till the end of the MS. At central H exhaustion, the moderate inflation of the radius brings the surface back to sub-critical velocities. 
%-----------------------------------------------------------------------------------------------------------------------------------------
\subsection{Mass loss and surface abundances \label{smdot}}

In \citet{mem06}, we show how rotating low-metallicity stars may lose more mass than expected mainly through two processes: first the reaching of the break-up velocity, which leads to a mechanical mass loss of moderate amplitude through a decretion disc; second the enrichment of the surface because of rotational mixing, which leads to a very large enhancement of the radiative mass loss. Now what happens if the metallicity is as low as strictly zero? The total mass lost at the end of MS remains modest in the present models (Table~\ref{tmlost}): between 0.03 $M_{\sun}$ for the 15 $M_{\sun}$ model (which reaches the critical limit only at the very end of the MS, when $X_\mathrm{c}=0.02$), and 16.4 $M_{\sun}$ for the 200 $M_{\sun}$ model (which reaches the break-up limit at $X_\mathrm{c}=0.43$). The mechanical mass loss removes only the outermost layers, which have a very low density, so only a little mass is removed during this phase.

\begin{figure*}
   \centering
    \resizebox{.73\hsize}{!}{\includegraphics{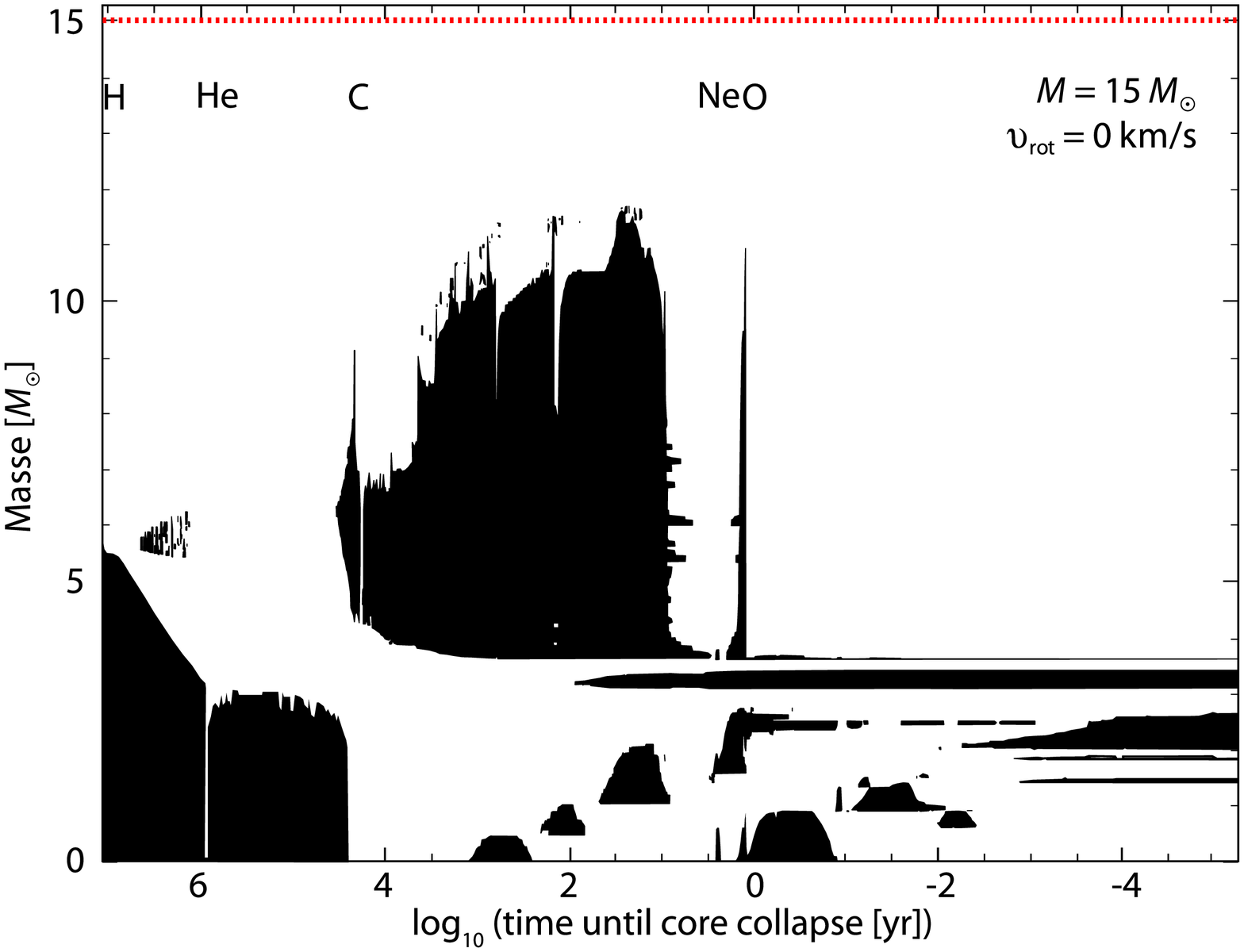}\hspace{1cm}\includegraphics{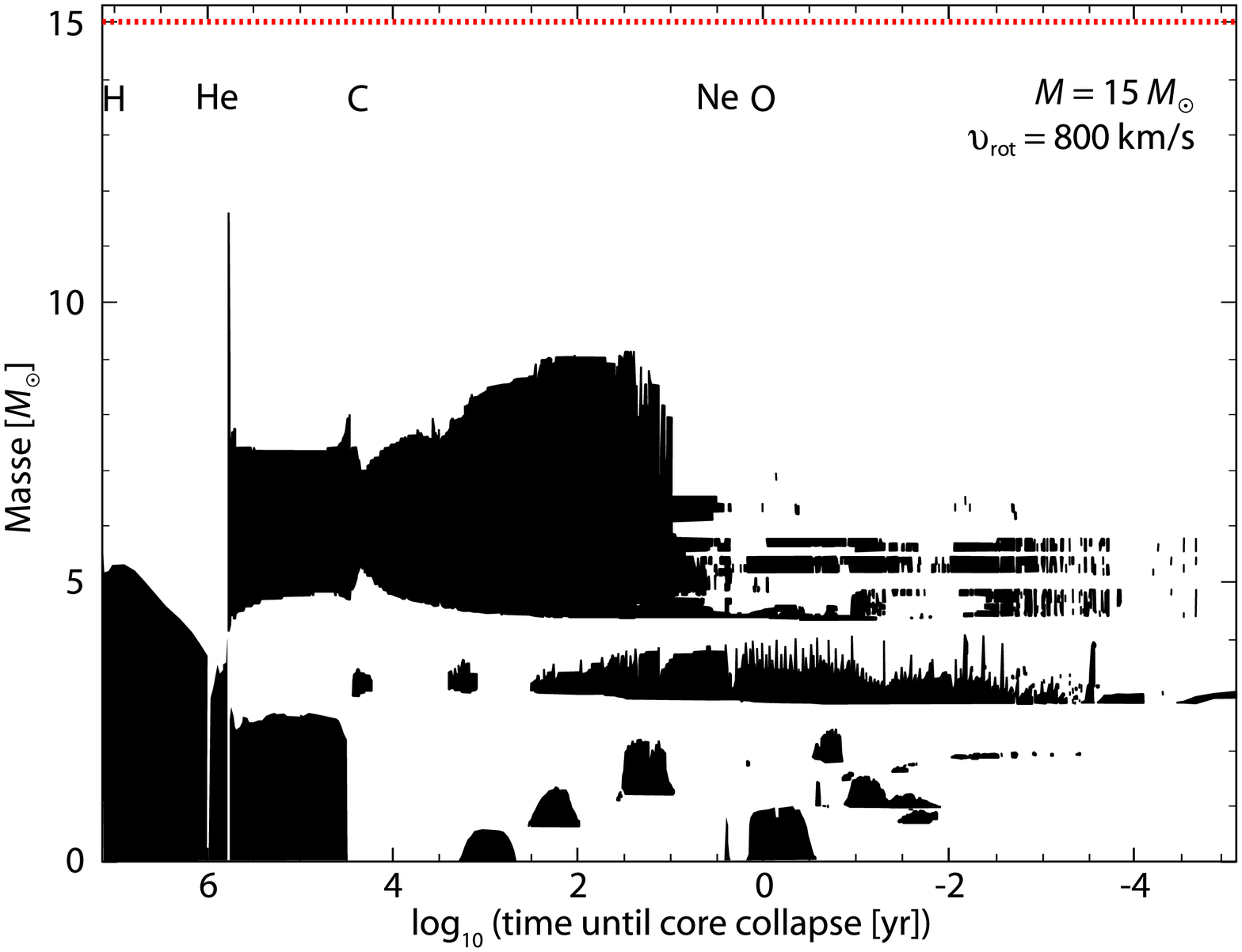}}
    \resizebox{.73\hsize}{!}{\includegraphics{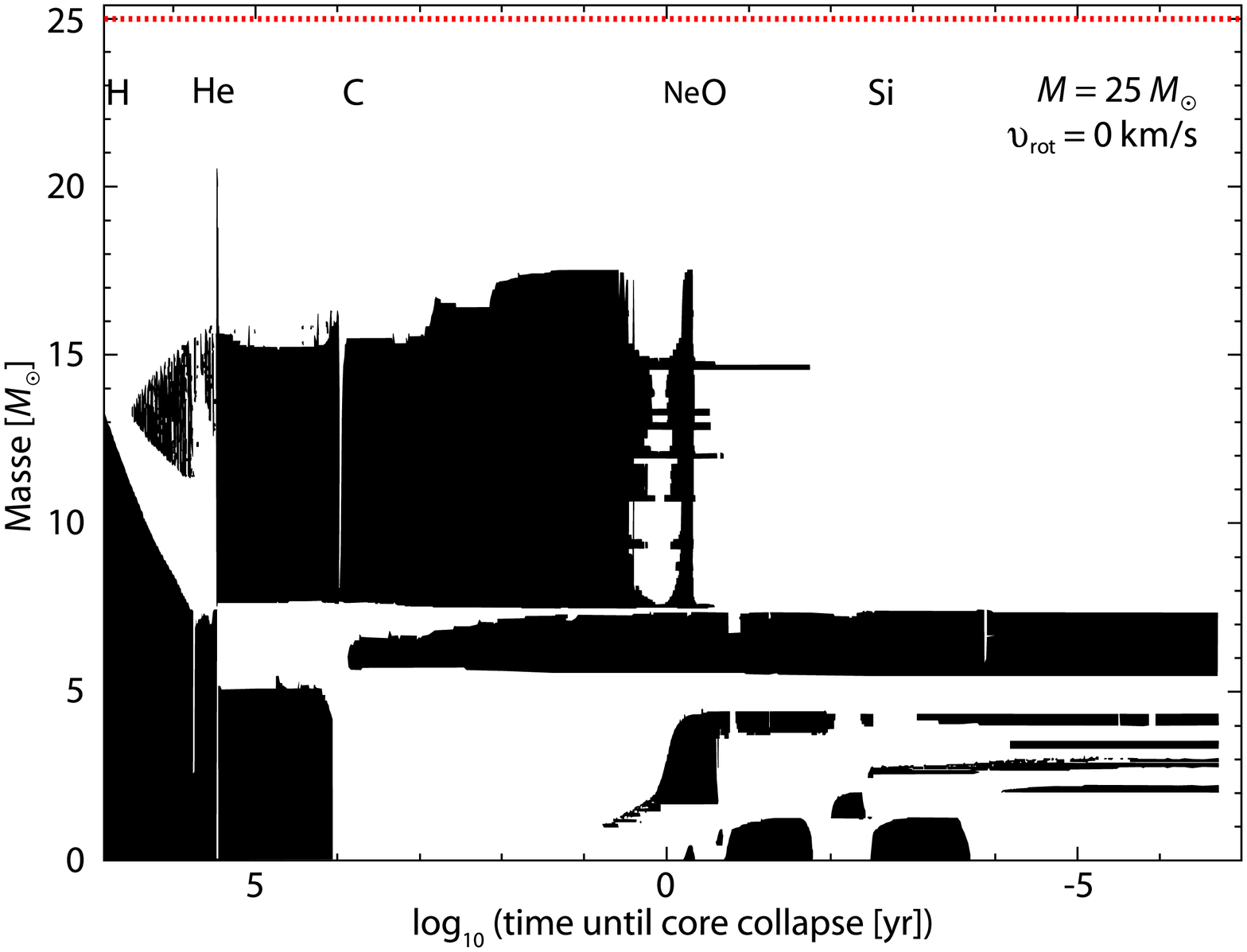}\hspace{1cm}\includegraphics{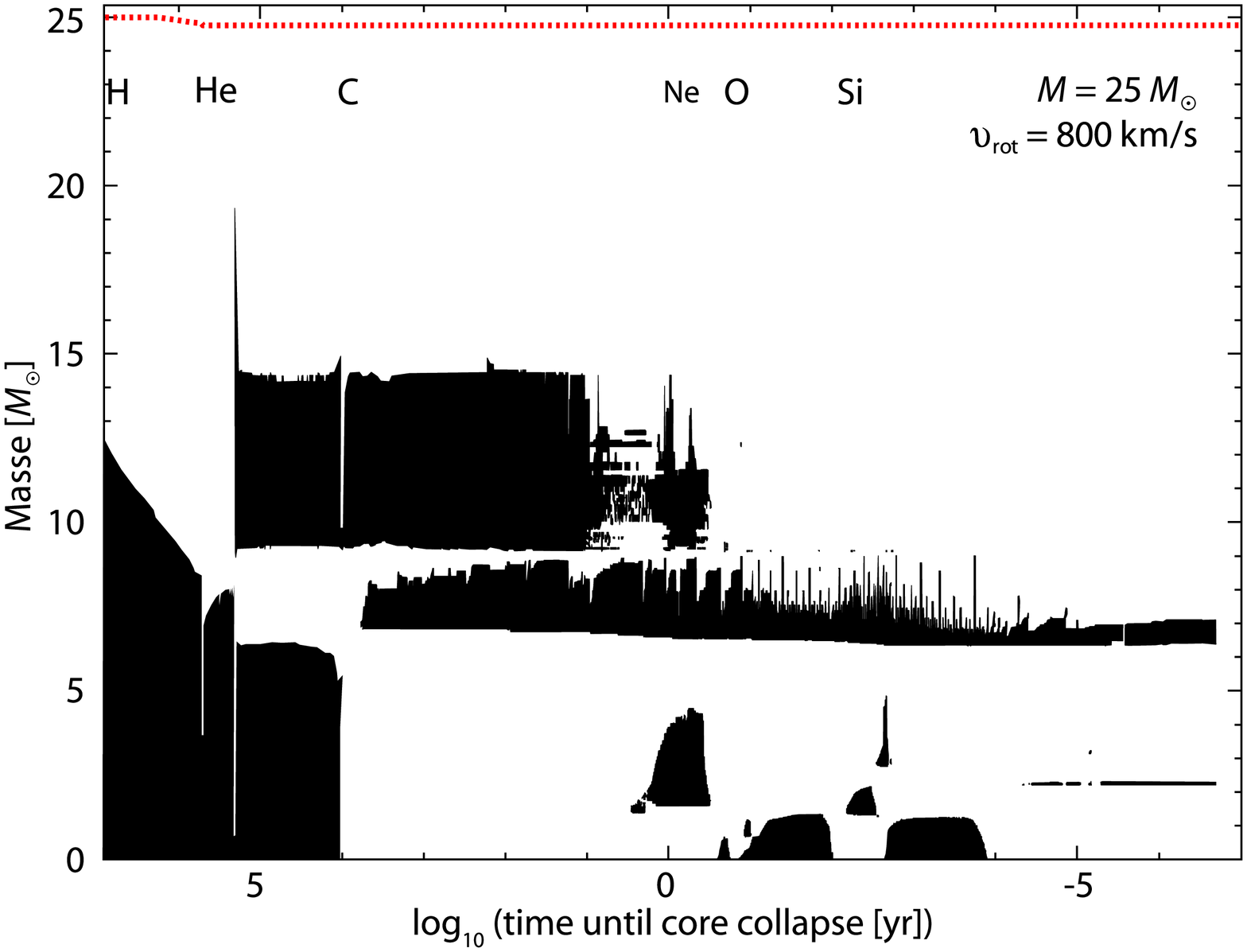}}
    \resizebox{.73\hsize}{!}{\includegraphics{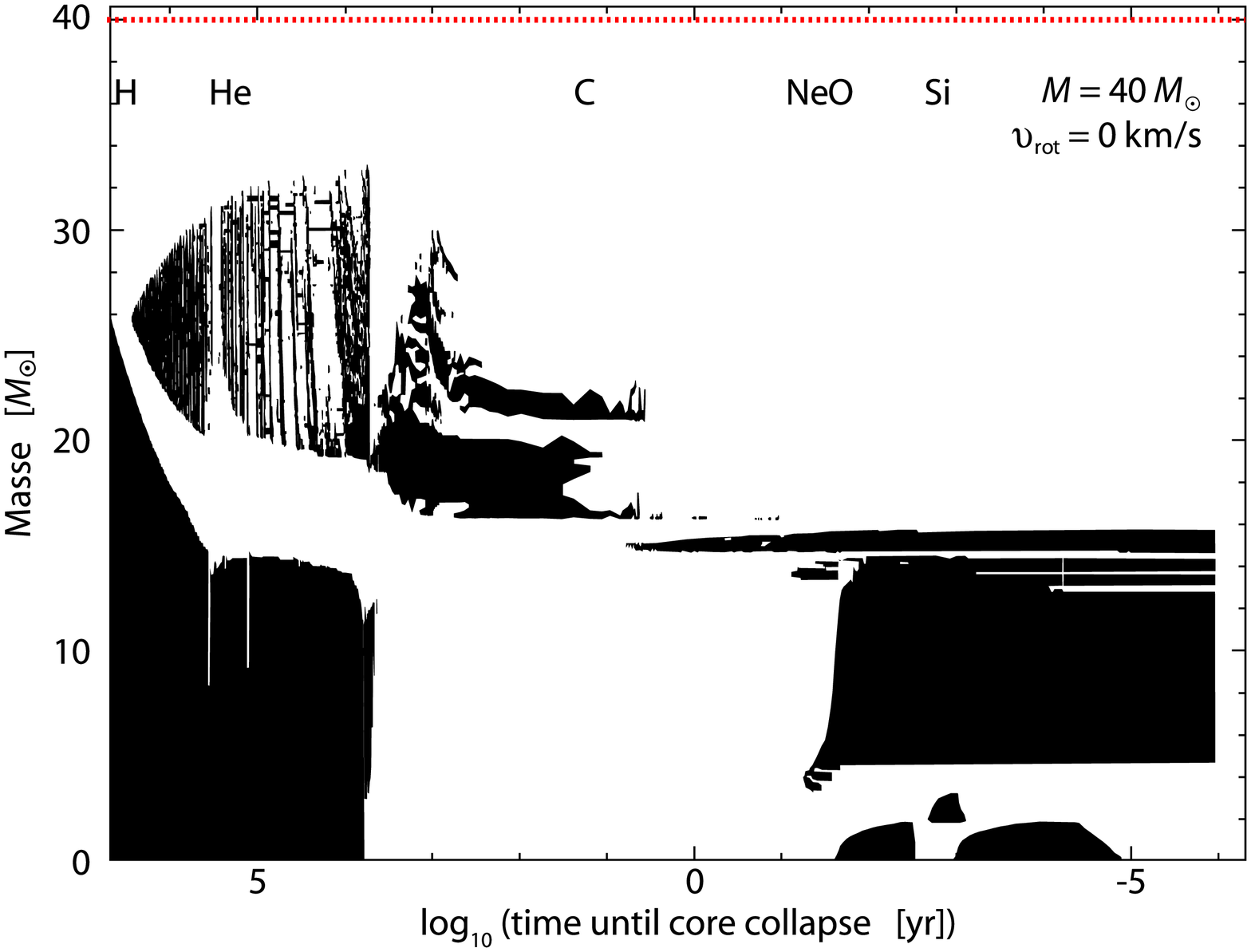}\hspace{1cm}\includegraphics{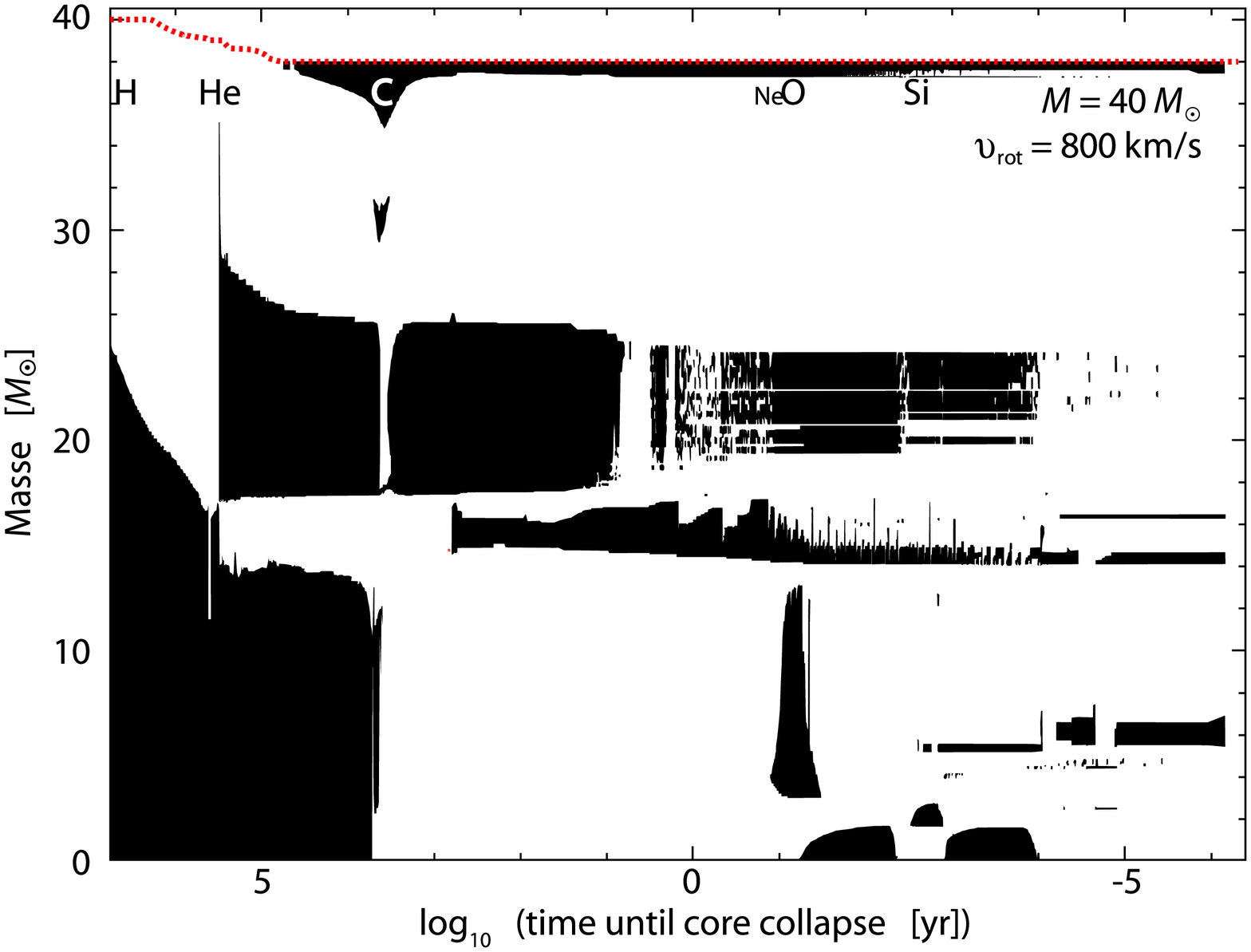}}
    \resizebox{.73\hsize}{!}{\includegraphics{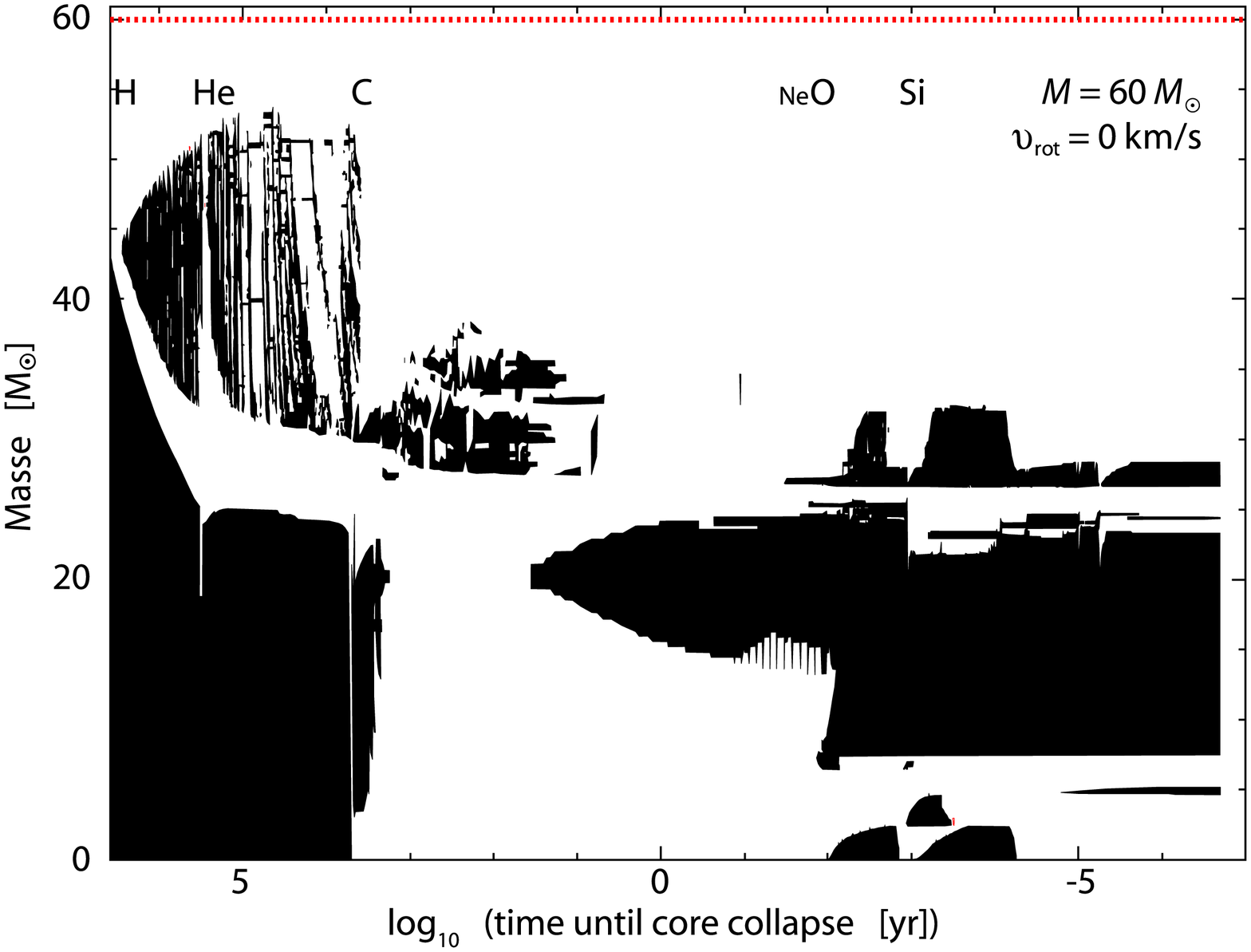}\hspace{1cm}\includegraphics{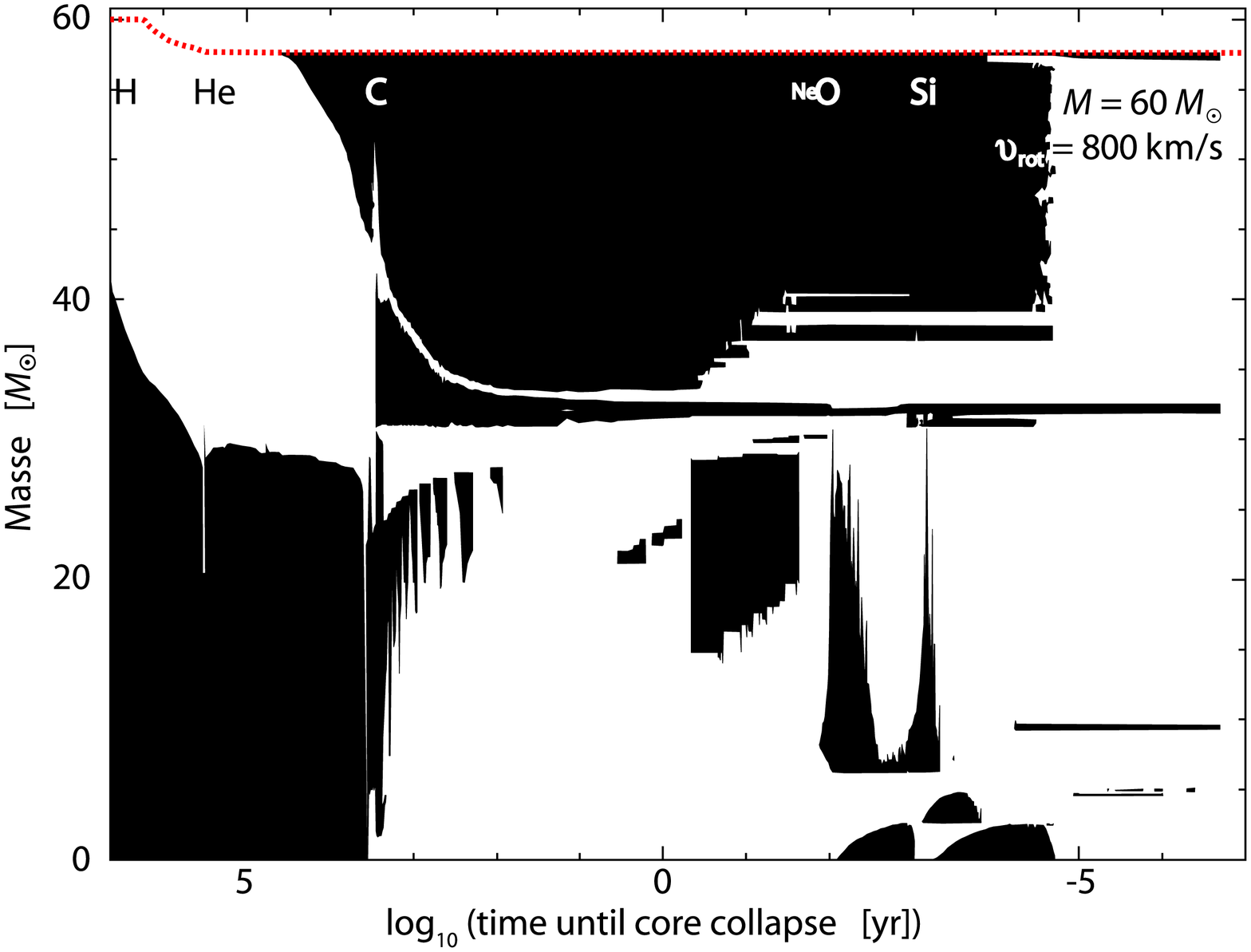}}
    \resizebox{.73\hsize}{!}{\includegraphics{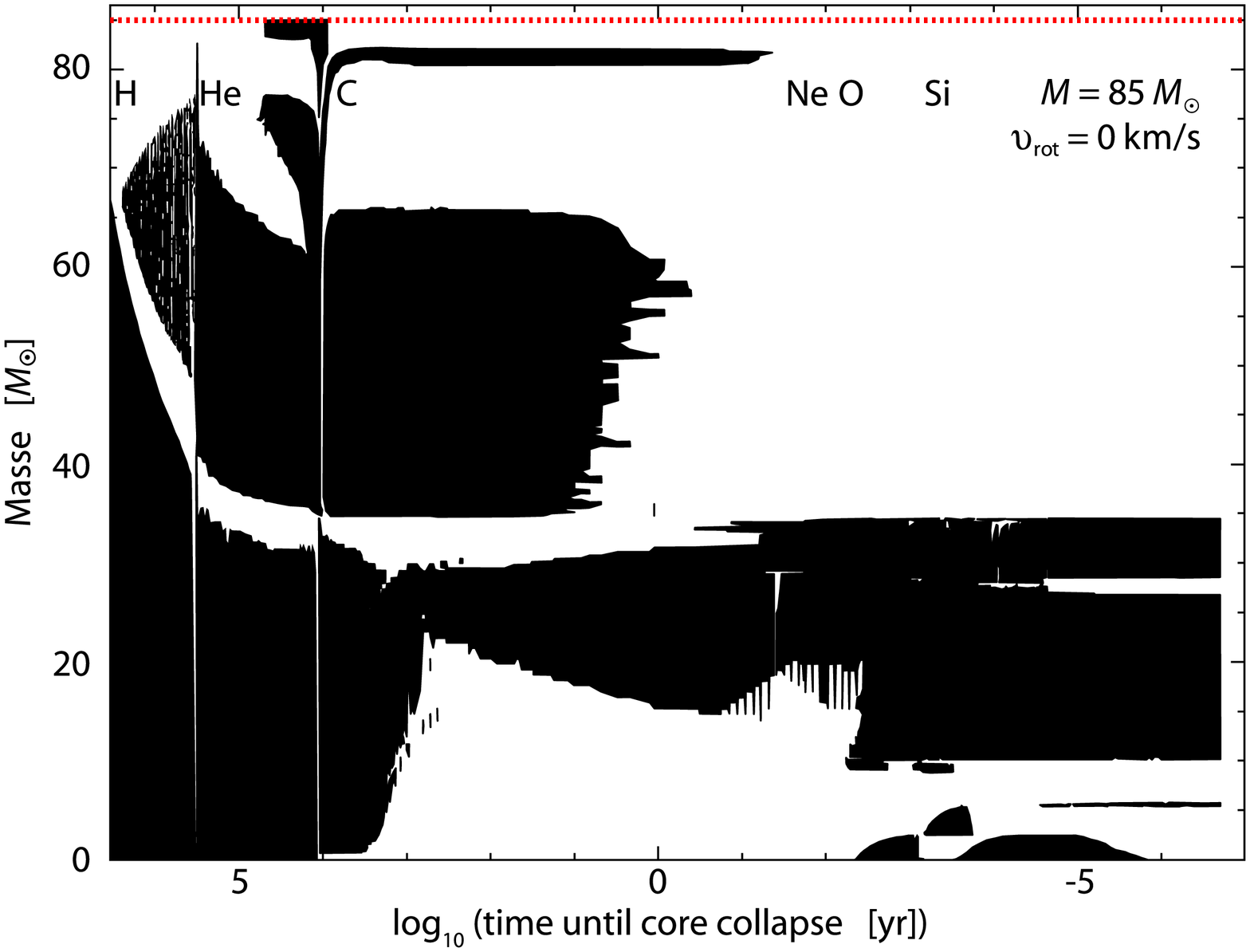}\hspace{1cm}\includegraphics{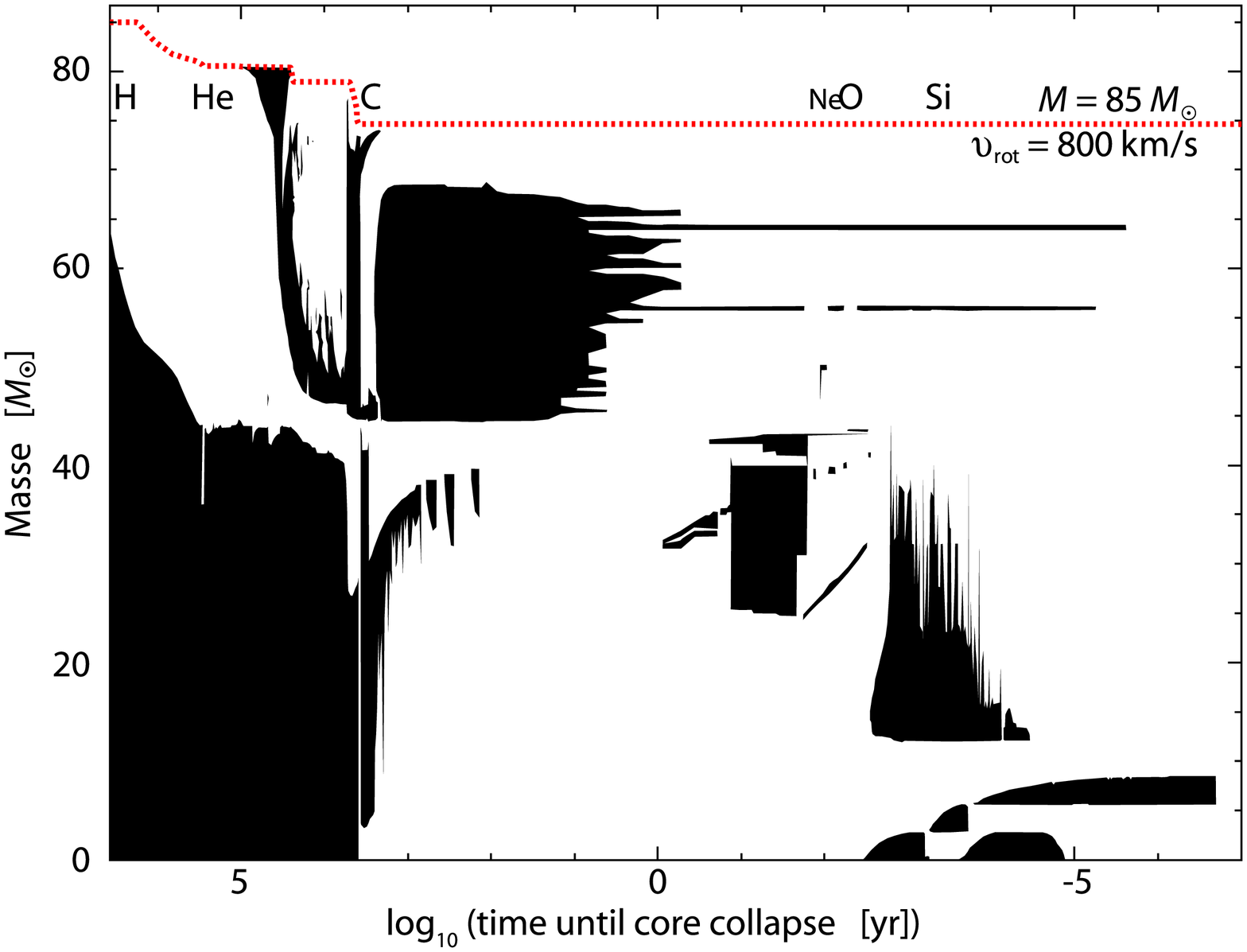}}
      \caption{Kippenhahn diagrams (evolution of the structure) for the models between 15 and 85 $M_{\sun}$. Left column: without rotation; right column: with rotation. The black areas show the convective zones. The (red) dotted line at the top of the figures is the total mass of the model. The on-set of the different burning phases is indicated at the top of each window.}
         \label{fkipp}
   \end{figure*}

After the MS, as mentioned in Sect.~\ref{sdhr}, the models remain in the blue part of the HRD most of the CHeB phase and then their redward incursion is moderate. The outer convective zone that develops remains very thin, so the dredge-up does not dig deeply enough (see the Kippenhahn diagrams in Fig.~\ref{fkipp}). Most of the post-MS evolution occurs with a surface metallicity much lower than $10^{-6}$. As a result, the mass lost during CHeB remains very low: it amounts to at most 6 $M_{\sun}$ in the case of the 85 $M_{\sun}$, which is the model that loses the most mass at this stage.
%:   fig fkipp

The total mass lost by the models during the whole evolution is small. The 200 $M_{\sun}$ model loses 20.7 $M_{\sun}$, the 85 $M_{\sun}$ model 10.4 $M_{\sun}$. The models have to be as massive as 40 $M_{\sun}$ or above to lose more than 1 $M_{\sun}$. This picture is very different from the evolution with just a small fraction of metals: the 60 $M_{\sun}$ at $Z=10^{-8}$ presented in \citet{mem06} loses 36 $M_{\sun}$, among which 27 $M_{\sun}$ are due to the enhancement of the radiative winds by the surface self-enrichment. At $Z=0$, a similar model loses 2.4 $M_{\sun}$ only. To check if the choice of a higher initial equatorial velocity could enhance significantly the mass loss in $Z=0$ models, we have computed a test model of 40 $M_{\sun}$ with $\upsilon_\mathrm{ini}=900$ km s$^{-1}$ (\textit{i.e.} $\Omega/\Omega_\mathrm{crit}=0.83$). We have evolved it until the end of CHeB. During the MS evolution, it reaches the critical limit when its central H mass fraction is $X_\mathrm{c}=0.43$ and loses 1.702 $M_{\sun}$ until the end of the MS. The total mass lost at the end of CHeB amounts to 1.734 $M_{\sun}$ which compares well with the 1.571 $M_{\sun}$ lost by the model with a slower $\upsilon_\mathrm{ini}$ at the same stage. We can thus conclude that the choice of a higher $\upsilon_\mathrm{ini}$ is not determinant to really change the situation about mass loss.

The non-rotating models do not lose mass at all: even the 200 $M_{\sun}$ model loses only a few hundredths of solar mass. We will see in Sect.~\ref{sfate} that some of our models may be able to contribute to the chemical enrichment of the medium only through their wind ejecta. In that context, while the non-rotating models do not contribute at all, the rotating ones will leave a modest imprint on their surrounding.

We cannot compare directly our results with the literature: the only study that does not compute the evolution at constant mass is the one by \citet{mar03} and their masses do not match exactly ours. Nevertheless, we can make a qualitative comparison. Our rotating 200 $M_{\sun}$ model experiences a much stronger mass loss than what they found for their rotating 250 $M_{\sun}$ model, which loses 3.35 $M_{\sun}$ only, that is 6 times less than our slightly less heavy model. Their models have been computed with a lower initial velocity, and despite the maximal coupling brought by the rigid body rotation, they attain the critical limit only at the very end of the MS phase. The mass loss they adopt then ($M_\mathrm{crit}= 10^{-3}\ M_{\sun} \mathrm{yr}^{-1}$) is about 2 orders of magnitude higher than that resulting from our treatment of the critical mass loss, but the period during which they apply it is short.
%==============================================================================
\section{Final structure and light elements yields \label{sfinstruc}}
%-----------------------------------------------------------------------------------------------------------------------------------------
\subsection{Kippenhahn diagrams \label{skipp}}

%:   tab tyields
\begin{table*}
\caption{Total stellar yields at the pre-supernova stage, in $M_{\sun}$. In parenthesis, the ``wind-only'' contribution is given for the models for which no SN contribution is expected (see Sect.~\ref{sfate}). The masses of the models are in $M_{\sun}$ and the velocities in km s$^{-1}$}
\label{tyields}
\centering
\begin{tabular}{r r c c c c c c c c c c}     % 12 columns
\hline\hline
$M_\mathrm{ini}$ & $\upsilon_\mathrm{ini}$ & $\frac{^{3}\mathrm{He}}{10^{-4}}$ & $^{4}$He & $^{12}$C & $^{13}$C & $^{14}$N & $^{16}$O & $^{17}$O & $^{18}$O & $^{22}$Ne & $Z$ \\
\hline \small
   9 &    0 & -0.53 & 1.17 & 0.19 & 5.54e-11 & 2.68e-8 & 0.24 & 2.74e-11 & 4.45e-9 & 1.58e-7 & 0.43 \\
   9 & 500& -0.65& 1.59 & 0.24 & 3.27e-5 & 1.81e-3 & 0.17 & 2.20e-7 & 8.20e-6 & 8.98e-6 & 0.41 \\
 15 &   0 & -1.02 & 1.24 & 0.22 & 5.94e-8 & 5.13e-6 & 0.86 & 1.93e-10 & 2.68e-9 & 8.73e-7 & 2.53 \\
 15 & 800& -1.23& 2.10 & 0.61 & 9.21e-4 & 4.99e-3 & 1.05 & 8.88e-7 & 7.17e-6 & 5.44e-4 & 2.10 \\
 25 &    0 & -1.97 & 2.32 & 1.26 & 2.09e-4 & 5.69e-3 & 2.78& 1.86e-6 & 1.79e-9 & 3.27e-6 & 4.93 \\
 25 & 800& -2.13 & 2.57 & 1.32 & 3.21e-5 & 1.08e-3 & 3.76& 2.75e-7 & 7.78e-6 & 5.54e-4 & 5.75 \\
  & & (-0.02) & (0.003) & (5e-14) & (1.59e-14) & (6.86e-12) & (2e-13) & (1.51e-15) & (6.51e-19) & (1.01e-26) & (7e-12) \\
 40 &    0 & -3.21 & 2.94 & 0.23 & 4.85e-8 & 3.09e-7 & 7.25 & 2.29e-9 & 1.55e-8 & 3.63e-6 & 10.2 \\
  & & (0.00) & (0.00) & (0.00) & (0.00) & (0.00) & (0.00) & (0.00) & (0.00) & (0.00) & (0.00) \\
 40 & 800& -3.32 & 5.32 & 1.63 & 8.41e-4 & 1.00e-1 & 7.57 & 4.04e-6 & 6.74e-5 & 3.30e-2 & 11.4 \\
  & & (-0.17) & (0.05) & (3e-8) & (7.96e-9) & (1.07e-6) & (4e-08) & (5.89e-11) & (5.43e-14) & (2.64e-11) & (1e-6) \\
 60 &    0 & -4.87 & 4.58 & 0.18& 9.07e-7 & 2.58e-6 & 13.6 & 1.69e-8 & 2.26e-7 & 1.89e-6 & 17.7 \\
  & & (0.00) & (0.00) & (0.00) & (0.00) & (0.00) & (0.00) & (0.00) & (0.00) & (0.00) & (0.00) \\
 60 & 800& -4.76 & 3.33& 0.79 & 2.39e-4 & 1.20e-2 & 16.9 & 1.89e-5 & 1.33e-4 & 5.24e-5 & 21.7 \\
  & & (-0.21) & (0.04) & (2e-8) & (7.10e-9) & (9.00e-7) & (9e-8) & (1.94e-10) & (8.18e-14) & (4.12e-11) & (1e-6) \\
 85 &    0 & -7.22 & 11.70& 0.28 & 1.46e-2 & 9.87e-1 & 19.7 & 6.63e-5 & 1.17e-7 & 1.18e-3 & 25.5 \\
 &  & (-0.004) & (0.01) & (3e-5) & (8.20e-6) & (2.42e-4) & (8e-6) & (7.26e-9) & (1.50e-11) & (3.99e-7) & (3e-4) \\
 85 & 800 & -6.15 & 5.31& 1.00& 6.44e-3 & 1.40e-1 & 26.2 & 1.24e-4 & 2.32e-7 & 6.47e-6 & 31.6 \\
\hline
\end{tabular}
\end{table*}

In Fig.~\ref{fkipp} we present the Kippenhahn diagrams (evolution of the structure) of the models between 15 and 85 $M_{\sun}$. Generally, rotation suppresses the convective instability outside the H-burning core by smoothing the abundance gradient of helium above the core. The CNO shell boost is clearly visible in the diagrams of the non-rotating 25 $M_{\sun}$ and the rotating 15, 25 and 40 $M_{\sun}$ models: the H-burning shell becomes suddenly convective and the core is abruptly reduced. We can discover in the rotating 60 $M_{\sun}$ diagram a similar boost, but only at the beginning of shell He burning. This boost left no signature in the HRD.

As mentioned above, the rotating 15 and 25 $M_{\sun}$ models remain in the blue part of the HRD and do not develop any outer convective zone (OCZ). The rotating 40 $M_{\sun}$ presents a very thin one and the 85 $M_{\sun}$ a very early and short-lived one. On the contrary, the 60 $M_{\sun}$ develops a deeper OCZ at the end of CHeB and for almost the rest of its evolution until the pre-SN stage. Its surface enrichment is thus larger than that presented by the 85 $M_{\sun}$ model. The only non-rotating model that develops an OCZ is the 85 $M_{\sun}$. Though this convective zone disappears quickly, it suffices to dredge up heavy elements brought near the surface by the CNO shell boost, and it is the model that shows the highest level of enrichment at the end of its evolution (see Table~\ref{tmlost}).

\citet{lsc00} show that at standard metallicity, the transition from convective to radiative core C burning occurs between 15 and 20 $M_{\sun}$. From the Fig.~1 of \citet{lcs01}, we deduce that at $Z=0$, this transition occurs at slightly higher masses, between 20 and 25 $M_{\sun}$. We note that our 15 $M_{\sun}$ models have convective C-burning cores, while cores are radiative in the case of our 25 $M_{\sun}$ models, in agreement with \citet{lcs01}. In the non-rotating models with $M \geq 40\ M_{\sun}$, the shell C burning is convective up to the end of the hydrostatic evolution, while in the rotating models, the shell experiences a few flashes but otherwise remains radiative.
%-----------------------------------------------------------------------------------------------------------------------------------------
\subsection{Stellar yields \label{syields}}

The yields calculated from stellar models are strongly correlated to the value of the CO-core mass, $M_\mathrm{CO}$. We give the values of $M_\mathrm{CO}$ of our models in Table~\ref{ttaucores}. They were determined as in \citet{hlw00} as the mass coordinate at which the abundance of He drops below $10^{-3}$ (in mass fraction). From these values, we determined the remnant masses of our models, $M_\mathrm{rem}$, with the relation between $M_\mathrm{rem}$ and $M_\mathrm{CO}$ described in \citet{maed92}

In Table~\ref{tyields} we present the total stellar yields for the light elements that are not significantly modified by the explosion \citep{lc02}. The total yields are the sum of the pre-SN yields ($mp_{im}^{\mathrm {preSN}}$) and the winds yields ($mp_{im}^{\mathrm {winds}}$). The pre-SN contribution from a star of initial mass $m$ to the stellar yield of an element $i$ is: $$mp_{im}^{\mathrm {preSN}} = \int_{M_\mathrm{rem}}^{M_\mathrm{fin}} \left[ X_i (m_r) - X_i^0 \right] {\mathrm d} m_r $$ with $X_i^0$ the initial abundance (in mass fraction) of element $i$ ($X_i^0=0$ for all elements except He), and $X_i (m_r)$ its final abundance at the Lagrangian coordinate $m_r$. The winds contribution is: $$mp_{im}^{\mathrm {winds}} = \int_0^{\tau(m)} \dot{M}(m,t) \left[ X_i^S (m,t) - X_i^0 \right] {\mathrm d} t $$ where $\tau(m)$ is the final age of the star with initial mass $m$, $\dot{M}(m,t)$ the mass loss rate when the age of the star is $t$, and $X_i^S (m,t)$ the surface abundance (in mass fraction) of element $i$ at age $t$. Most of the time, the total stellar yields are dominated by the pre-SN contribution.
%:   fig fycomp0
   \begin{figure}
   \centering
    \resizebox{\hsize}{!}{\includegraphics{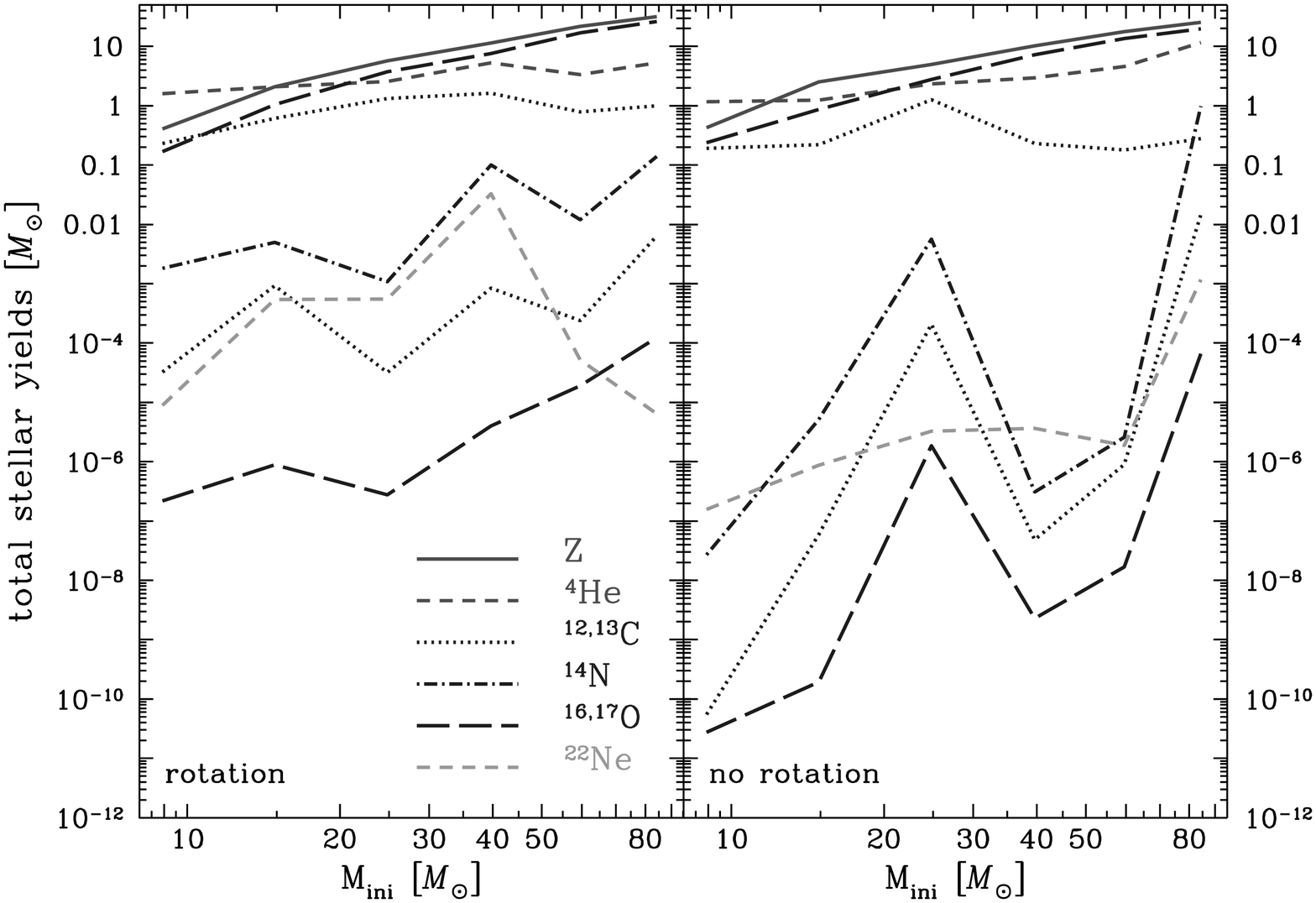}}
      \caption{Total stellar yields as a function of initial mass for our $Z=0$ models. \textit{Left:} models with rotation; \textit{right:} non-rotating models.}
         \label{fycomp0}
   \end{figure}

We did not include the 200 $M_{\sun}$ models in Table~\ref{tyields} because they have not undergone the shell He-burning phase, which modifies noticeably the results. The 9 $M_{\sun}$ models were also stopped before central C burning, but as we mentioned, the core was strongly degenerated at carbon ignition, which could lead to the complete destruction of the star without further processing. Note that the carbon ignition occurred at the edge of the core, so there is a possibility that the star survived C burning. In that case, since the mass included below the He-burning shell is smaller than the limiting mass for Ne burning (1.37 $M_{\sun}$), the star is expected to collapse in a Type II SN by electrons capture on a degenerate O-Ne-Mg core \citep{nomoto84}. This would modify the pre-SN yields presented here since the C burning would have processed the matter of the core. We would thus expect a reduction of $^{12}$C and an enhancement of $^{13}$C, $^{14}$N, $^{16}$O and $^{22}$Ne. In the case of the 15 $M_{\sun}$ models, the evolution has been stopped at the end of central O burning. At that time, the mass interior to the C-burning shell is higher than Chandrasekhar mass, so the star is expected to undergo an off-centre Si flash and to end with an Fe core \citep{nomoto84}. We then expect the yields presented here to be modified slightly (reduction of C, N and Ne, enhancement of O). We have checked the amplitude of the variation when the yields are calculated at the end of O burning or at the end of Si burning in the models that have been followed until the pre-SN phase, so we do not expect  the yields of the 15 $M_{\sun}$ models given in Table~\ref{tyields} to vary by more than 15\%.

Rotation leads to larger yields for almost all the isotopes  and for almost all the masses (see Fig.~\ref{fycomp0}). The exceptions are the non-rotating models that undergo a strong CNO shell boost: the 25 and 85 $M_{\sun}$, which produce larger amounts of $^{13}$C and $^{14}$N. The 25 $M_{\sun}$ overtakes its rotating counterpart also in the production of $^{17}$O, and the 85 $M_{\sun}$ produces also more $^4$He and $^{22}$Ne than the rotating one. Otherwise, the enhancement of the yields brought by the rotational mixing may reach orders of magnitude, especially so for the production of primary $^{14}$N: a factor $10^5$ between the rotating and non-rotating 9 $M_{\sun}$ models and even a factor $10^6$ in the case of the 40 $M_{\sun}$ models. We note that rotation makes the production of $^{14}$N robust through the whole mass range, while there are enormous differences from mass to mass in the non-rotating models.
%:   fig fycomp
   \begin{figure*}
   \centering
    \resizebox{\hsize}{!}{\includegraphics{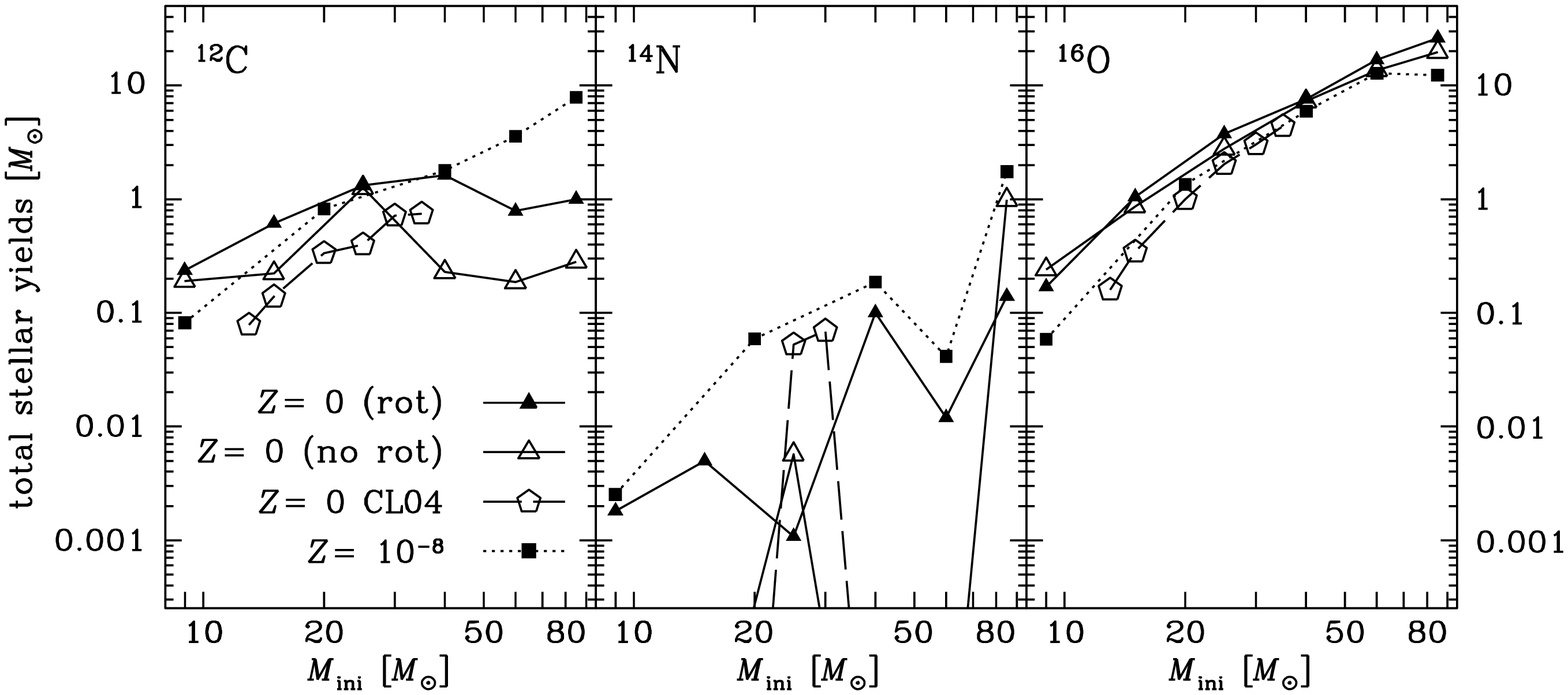}}
      \caption{Yields comparison between the non-rotating $Z=0$ models from \citet[CL04, open pentagons]{chieflim04}, the rotating $Z=10^{-8}$ models from \citet[filled squares]{hir07} and our rotating (filled triangles) and non-rotating (open triangles) $Z=0$ models. \textit{Left:} $^{12}$C; \textit{centre:} $^{14}$N; \textit{right:} $^{16}$O.}
         \label{fycomp}
   \end{figure*}

Figure~\ref{fycomp} shows a comparison of the CNO yields of all our $Z=0$ models with the yields of the non-rotating models at $Z=0$ from \citet{chieflim04} and those of the rotating models at $Z=10^{-8}$ from \citet{hir07}. For $^{12}$C (\textit{left}) and $^{16}$O (\textit{right}), our non-rotating 15 and 25 $M_{\sun }$ yields compare within a factor 2 or 3 with those of \citet{chieflim04}. Their 25 $M_{\sun}$ produces also a huge amount of $^{14}$N (\textit{centre}), even a factor $\sim$ 10 higher than ours, so the CNO boost of the H-burning shell seems to be a robust feature in this mass domain. Compared to the $^{12}$C produced at $Z=10^{-8}$, the production of our rotating 60 and 85 $M_{\sun }$ models is smaller, while it is almost equal or larger for the 9 to 40 $M_{\sun }$ models. This can be understood because our high mass models do not undergo a shell boost, so their He-burning core is larger, which leads to a lower $^{12}$C yield. The $^{16}$O production is about the same at $Z=10^{-8}$ and $Z=0$. Our PopIII models have slightly larger  $^{16}$O yields than the models of \citet{hir07}, but we use a higher overshooting parameter which leads to larger CO cores and thus larger $^{16}$O production.

An interesting point is the production of primary nitrogen at very low metallicity. \citet{mm8} show that this production increases when the metallicity decreases, and \citet{hir07} confirms the trend. We should thus expect a still higher production at $Z=0$, but this is not the case. $^{14}$N is systematically less produced at $Z=0$ than at $Z=10^{-8}$. We relate this change of behaviour to the fact that the $Z=0$ stars burn their hydrogen at temperatures hot enough to burn some He as well and do not evolve into the red part of the HR diagram (or do so very slowly during CHeB) as can be seen in Fig.~\ref{fdhr} (\textit{left}). Due to this fact, the stars undergo much less mixing during the CHeB phase. Indeed, no redwards evolution means that the gradient of the angular velocity at the border of the He-burning core remains much shallower and triggers much less mixing. On the contrary, when a small amount of metals is present, stars begin to burn their hydrogen through the CNO cycle from the start. Thus the central temperatures during the MS phases are well below those required for He-burning. At the end of the MS phase, the core contracts and the stars evolve quickly to the red. A steep angular velocity gradient appears at the core border and triggers strong mixing. Figure~\ref{fycomp} shows that there is even a difference  of almost two orders of magnitude between the production of the 20 $M_{\sun}$ at $Z=10^{-8}$ and the 25 $M_{\sun}$ at $Z=0$. This can be explained as follows: the $Z=10^{-8}$ model experiences a second CNO shell boost during shell He burning, which enhances its $^{14}$N production by about two orders of magnitude.
%-----------------------------------------------------------------------------------------------------------------------------------------
\subsection{Fate of the models \label{sfate}}

At high metallicity, massive stars experience a very strong mass loss. Their final mass is therefore very small \citep[even smaller than the final mass of stars with an initial mass $M_\mathrm{ini} \leq 25\ M_{\sun}$, see][]{maed92}. Thus, the fate of all massive stars is a SN explosion, whatever their initial mass.

At very low metallicity, this is no longer true. \citet{HW02} and \citet{hfw03} determine the limiting $M_\mathrm{He}$ for which the fate will be:
\begin{itemize}
\item a Type II SN: $M_\mathrm{He} < 9\ M_{\sun}$,
\item a black hole (BH) by fallback: $9\ M_{\sun} \leq M_\mathrm{He} < 15\ M_{\sun}$,
\item a direct BH: $15\ M_{\sun} \leq M_\mathrm{He} < 40\ M_{\sun}$,
\item a pulsational pair-instability followed by a SN with BH formation: $40\ M_{\sun} \leq M_\mathrm{He} < 64\ M_{\sun}$,
\item a pair-instability SN (PISN): $64\ M_{\sun} \leq M_\mathrm{He} \leq 133\ M_{\sun}$,
\item or a direct BH collapse: $M_\mathrm{He} > 133\ M_{\sun}$.
\end{itemize}
Note that the limits above are obtained by non-rotating models, and they may not be appropriate to rotating models, as shown by \citet{fryer01}. But since we are lacking more accurate determinations, we applied them to all our models. A star that ends its life as a BH is supposed to contribute to the chemical enrichment of its surrounding only through its winds. We give in parenthesis in Table~\ref{tyields} this ``wind-only'' contribution for the models that, according to the limits determined above, end their life as BH by fallback or by direct collapse.

The winds yields are very small, since we have seen that the winds are very weak (Sect.~\ref{smdot}). The 25 $M_{\sun}$ for instance, releases a total metals amount of less than $10^{-7}\ M_{\sun}$ through its winds while it produces a little more than 5 $M_{\sun}$ in the SN scenario. The 60 $M_{\sun}$ contributes to the enrichment of the early Universe with a mere $10^{-5}\ M_{\sun}$ of metals. Note that the net enrichment depends of course on the amount of metals ejected, but also on the volume in which they are diluted. The wind contribution is expected to be a slow one, so it will be diluted in a smaller volume. The net enrichment may not be so much smaller compared to a SN contribution diluted in a larger volume.

%:   tab trotmfe
\begin{table}
\caption{Some rotational properties of the iron core of our models: the mass $M_\mathrm{Fe}$ where the abundance of the sum $^{48}$Cr + $^{52}$Fe + $^{56}$Ni amounts to 50\% in mass fraction; the angular velocity $\Omega$ taken in the middle of the iron core (the profile of $\Omega(m)$ is almost flat in the core, so this value represents the mean value of $\Omega$ in the iron core); the maximum specific angular momentum $j$ inside the core, corresponding to the first peak seen in Fig.~\ref{fj85} (\textit{left}); the total angular momentum $\mathcal L$ and the parameter $\beta=E_\mathrm{rot}/|E_\mathrm{pot}|$, both taken at the border of the iron core.}
\label{trotmfe}
\centering
\begin{tabular}{r c c c c c}
\hline\hline
    Mass & $M_\mathrm{Fe}$ & $\Omega$ & $j$ & $\mathcal L$ & $\beta$ \\
   $\left[M_{\sun}\right]$ & $\left[M_{\sun}\right]$ & [s$^{-1}$] & [cm$^2$ s$^{-1}$] & [g cm$^2$ s$^{-1}$] & \% \\
\hline
25 & 1.327 & 0.4232 & 4.073e+16 & 5.837e+49 & 1.5833 \\
40 & 1.555 & 0.3547 & 3.667e+16 & 5.963e+49 & 1.0046 \\
60 & 2.516 & 0.1578 & 4.338e+16 & 1.152e+50 & 0.5449 \\
85 & 2.779 & 0.1062 & 3.913e+16 & 1.182e+50 & 0.3624 \\
\hline
\end{tabular}
\end{table}
%:   fig fj85
   \begin{figure*}
   \centering
    \resizebox{\hsize}{!}{\includegraphics{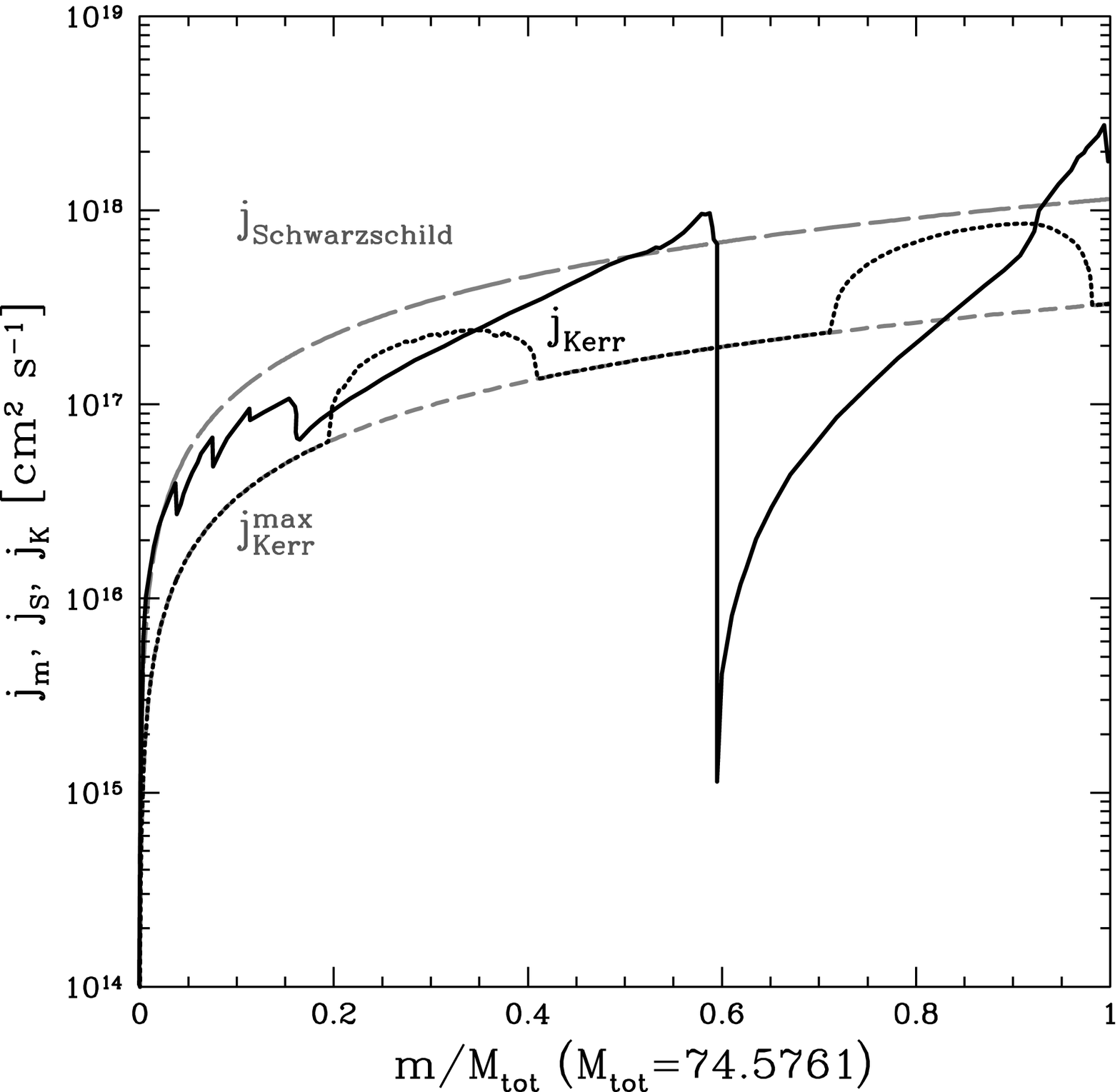}\hspace{1cm}\includegraphics{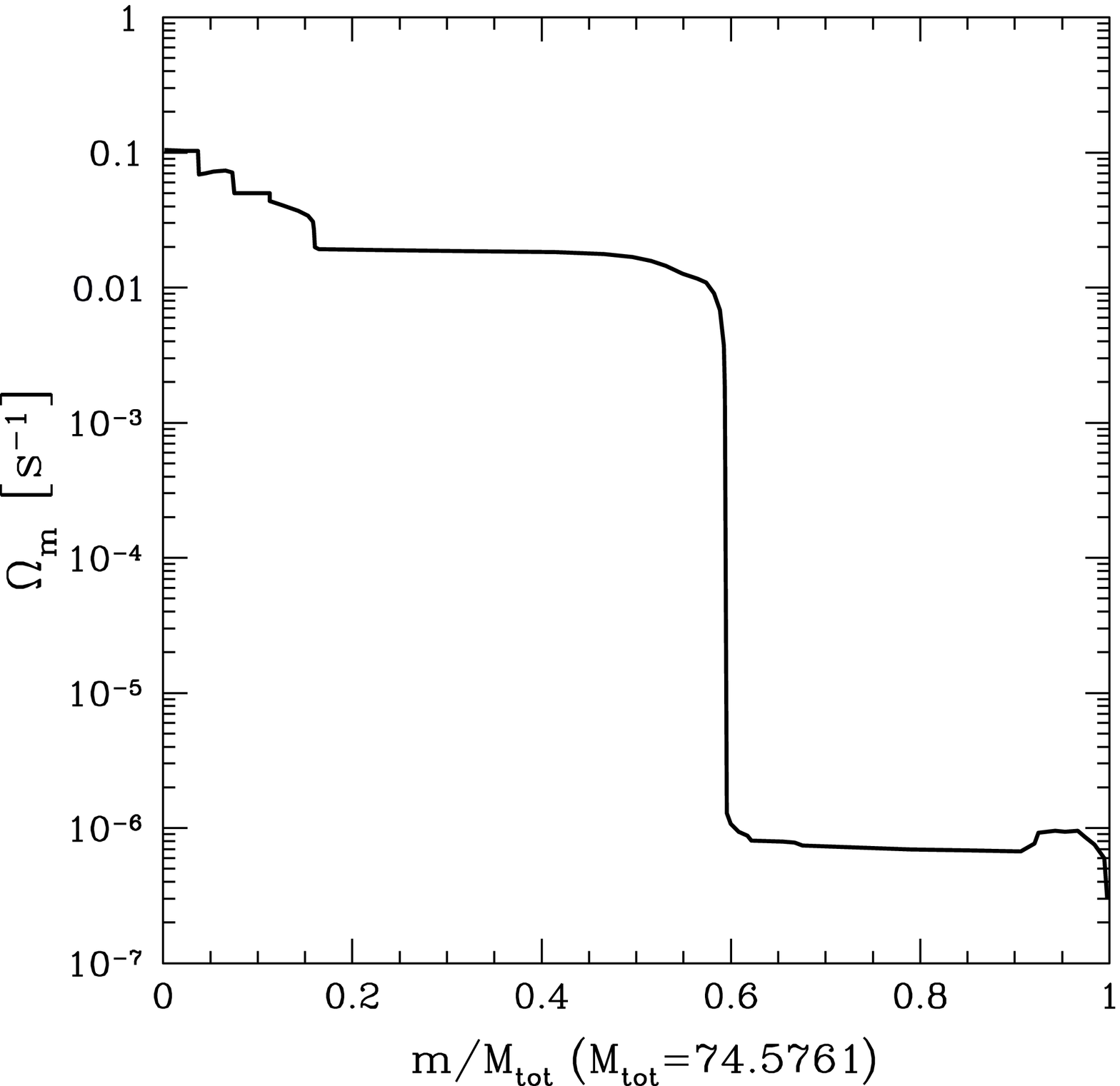}}
      \caption{\textit{Left:} profile of the specific angular momentum $j_\mathrm{m}$ inside our 85 $M_{\sun}$ at the end of hydrostatic core Si-burning (solid line). The dotted line is $j_{\rm K}=r_{\rm LSO}\,c$ \citep[ p. 428]{ST83}, where the radius of the last stable orbit, $r_{\rm LSO}$, is given by $r_{\rm ms}$ in formula (12.7.24) from \citet[p. 362]{ST83} for circular orbit in the Kerr metric. $j_{\rm K}$ is the minimum specific angular momentum necessary to form an accretion disc around a rotating black hole. $j_\mathrm{Schwarzschild}=\sqrt{12}Gm/c$ (long-dashed line) and $j_\mathrm{Kerr}^\mathrm{max}=Gm/c$ (short-dashed line) are the minimum specific angular momentum necessary for a non-rotating and a maximally-rotating black hole, respectively. \textit{Right:} profile of $\Omega_\mathrm{m}$ inside the same model.}
         \label{fj85}
   \end{figure*}

%:   fig fchemev
   \begin{figure*}
   \centering
    \resizebox{\hsize}{!}{\includegraphics[width=\textwidth]{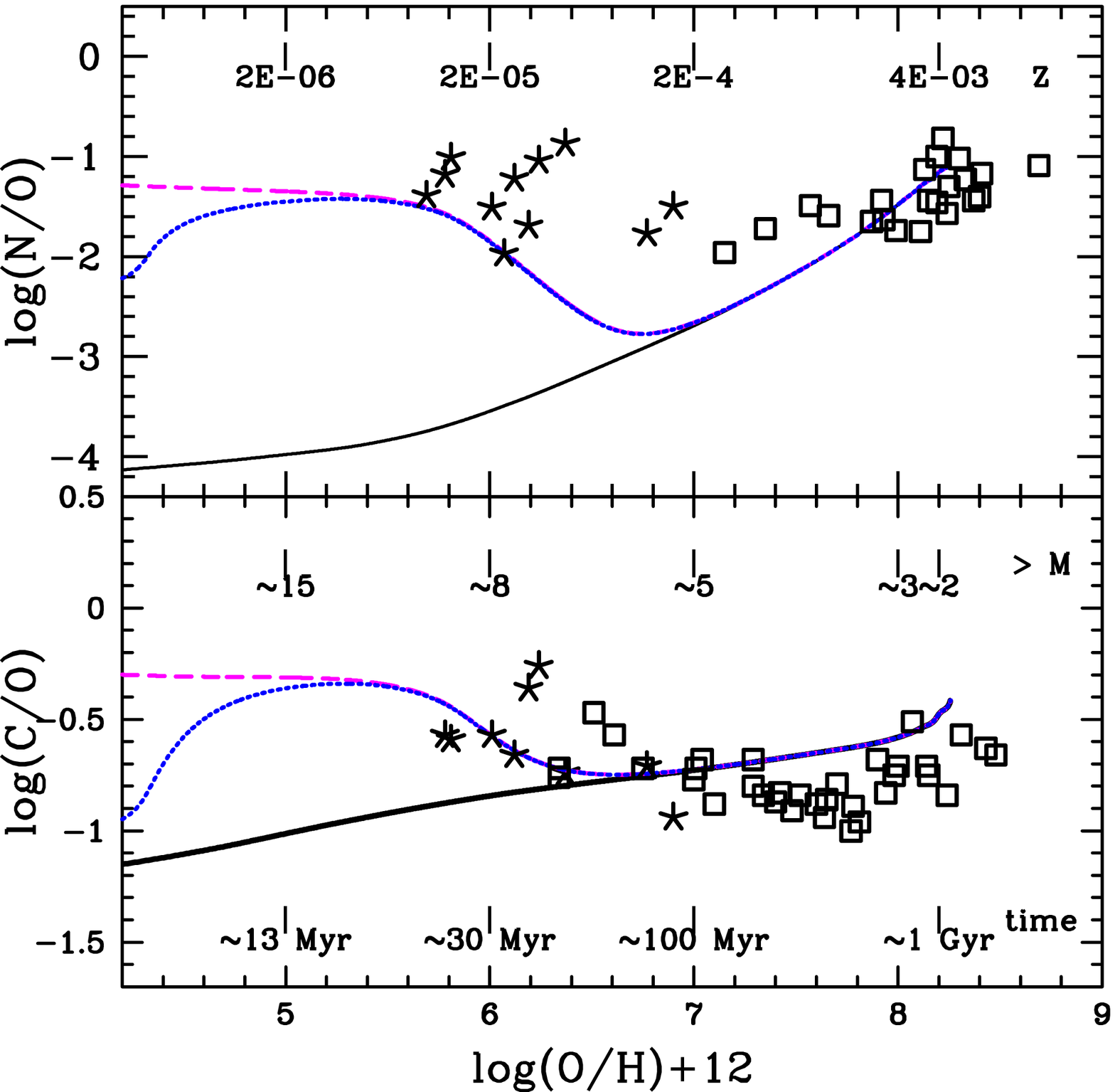}\hspace{1cm}\includegraphics[width=\textwidth]{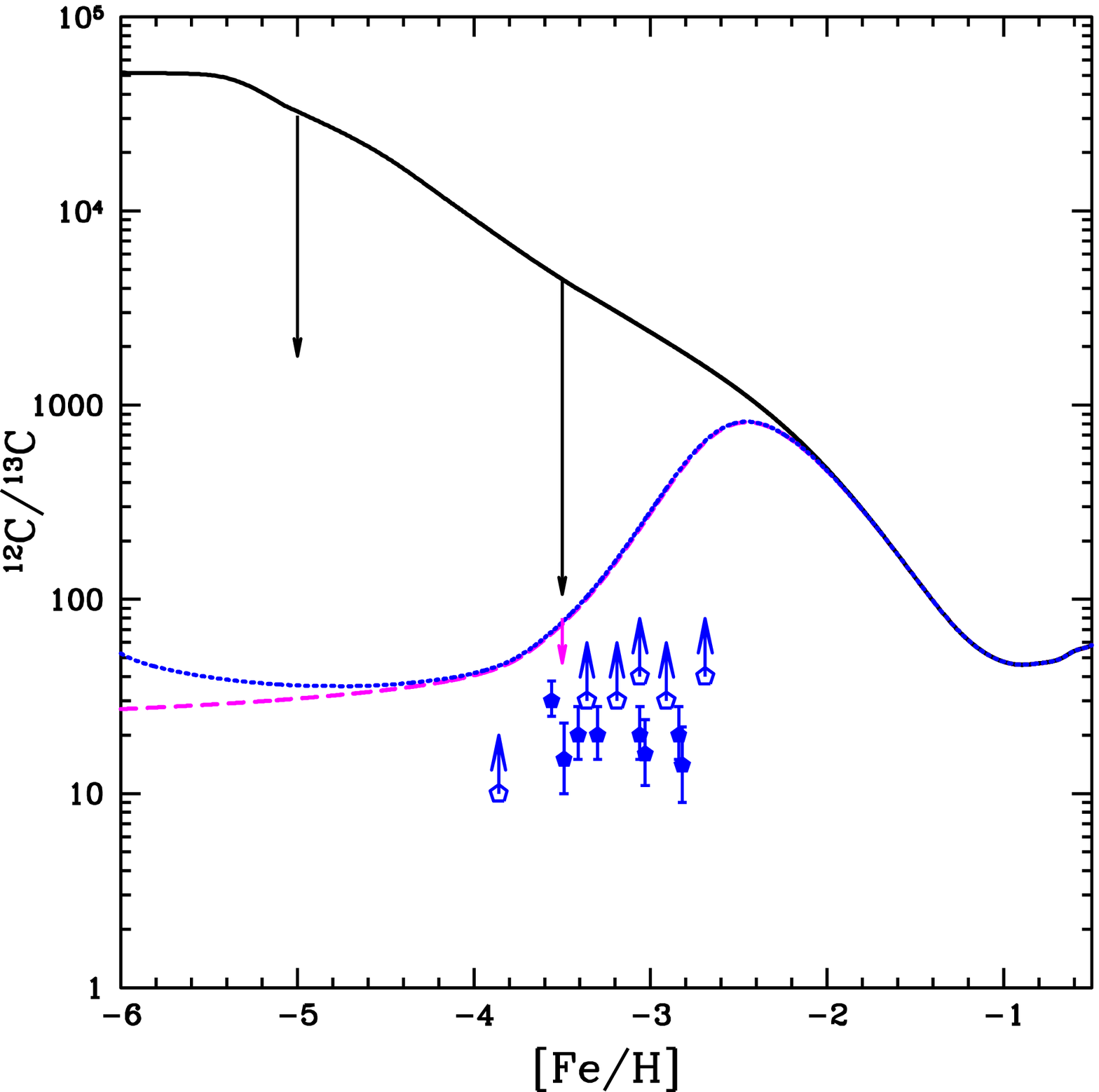}}
      \caption{The (black) solid curve is the CEM obtained with the stellar yields of slow rotating $Z=10^{-5}$ models from \citet{mm8} and \citet{hmm12}. The (magenta) dashed line includes the yields of fast rotating $Z=10^{-8}$ models from \citet{hir07} at very low metallicity. The (blue) dotted curve is obtained using the yields of the $Z=0$ models presented in this paper up to $Z=10^{-10}$. \textit{Left:} evolution of the N/O and C/O ratios. Data points are from \citet[open squares]{israel04} and \citet[stars]{spite05}. \textit{Right:} evolution of the $^{12}$C/$^{13}$C ratio. Data points are \textit{unmixed} stars from \citet{S06}: open pentagons are lower limits. The arrows going down from the theoretical curves indicate the final $^{12}$C/$^{13}$C observed in giants (after the dredge-up), starting from the initial composition values given by the stellar models \citep[see][]{chiap08}.}
         \label{fchemev}
   \end{figure*}

Rotation changes the mass of the He-cores at the end of the evolution. They are usually more massive because of the rotationally-induced mixing, and this may have consequences for the way the star ends its life. While the 25 $M_{\sun}$ without rotation ends as a Type II SN, the rotating one is supposed to produce a weak explosion and a BH by fallback. The non-rotating 85 $M_{\sun}$ remains in the direct BH domain while the rotating one reaches a He-core mass that implies pulsational pair-instability, followed by a Fe-core collapse into BH. Its true yields must thus be very close to the pre-SN value.

As mentioned in Sect.~\ref{sdhr}, none of our models meet the conditions for becoming WR stars. They end their evolution having kept their envelope, and thus seem to fail to become a GRB progenitor as defined in the collapsar model of \citet{woos93}. However, we have seen that the rotating 85 $M_{\sun}$ is likely to undergo pulsational pair instability\footnote{This process has not been followed in the present models.} and thus lose some mass in this process. If the mass lost is large, it may become a GRB progenitor since the very low core-envelope coupling has maintained a high angular momentum in the core. Fig.~\ref{fj85} (\textit{left}) shows that in the innermost parts of this model, the specific angular momentum is higher than that needed to create an accretion disc around a BH.

More generally, we expect that rotation will change the way a star collapses. \citet{shim01} show that an anisotropy of only 10\% in the neutrinos flux can boost the energy of the explosion. \citet{wald05} and \citet{yama05} show that the rotation-induced asphericity of the matter accretion rate is an even more important factor that helps producing a robust explosion. The geometry of the system leads most probably to jet-like explosion. Table~\ref{trotmfe} gives the rotational characteristics of the iron core of our models and Fig.~\ref{fj85} (\textit{right}) the distribution of angular velocity inside the 85 $M_{\sun}$ model. All models have kept a high total angular momentum, and meet the conditions to produce a maximally rotating BH. Though detailed calculations would be needed to ascertain this point, the true fate of our models could be much more explosive than previously supposed. In the chemical evolution model discussed below (see Sect.~\ref{sgalevol}), we shall use the total yields, implicitly assuming that the material above $M_\mathrm{rem}$ will be ejected.

The centrifugal force has a stabilising effect on the collapsing core, so the core bounce may occur at sub-nuclear densities. This alters the emission of gravitational waves (GW). \citet{ott04} show that rotation increases the amplitude of GW because of the induced anisotropy. But if rotation is really fast, for $\beta=E_\mathrm{rot}/|E_\mathrm{pot}| > 0.3\%$, the amplitude of GW becomes smaller, and the characteristic frequency is shifted to lower frequencies. Except for the 85 $M_{\sun}$, our models have a high $\beta$ so their GW emission is expected to be damped by their high rotation rate. \citet{suwa07} show that very massive collapsing PopIII stars may contribute to the GW background and be detectable by future facilities as DECIGO, the successor of LIGO. In that framework, our 85 $M_{\sun}$, with $\beta=0.3\%$, is expected to contribute to this background.
%==============================================================================
\section{Chemical evolution of the Galaxy with \boldmath $Z=0$ yields \label{sgalevol}}

Yields of extremely low-metallicity models are constrained by the most metal-poor stars observed in the halo of our Galaxy. They are supposed to trace the local chemical evolution of their surrounding medium at the time of their birth. The notion of time is important to keep in mind because stars with different masses have different lifetimes: the elements they produce are not released at the same moment. The oldest stars observed nowadays may have form when only massive stars had time to contribute to the enrichment of the medium. If we want to understand the abundances observed at the surface of these old stars, we cannot just integrate the yields obtained by a set of models with an IMF, because we would lose track of the time. What we need is a complete chemical evolution model (CEM).

Such a CEM is computed by \citet{chiap06} for example\footnote{The details of the models can be found in \citet{chiapnic9}, where they show that such a model reproduces nicely the metallicity distribution of the Galactic halo stars. This means that the timescale for the chemical enrichment in the halo is well constrained.}. They compare their predictions for the evolution of N/O and C/O upon the adoption of different sets of stellar yields with the surface ratios determined for the metal-poor stars observed by \citet{israel04} and \citet{spite05}. They show that the N/O ratio is much higher than what was predicted by a CEM using the yields of the slow-rotating $Z=10^{-5}$ models from \citet{mm8} down to $Z=0$. When adding the yields of the fast-rotating $Z=10^{-8}$ models from \citet{hir07}, the fit is much improved. The same improvement is found for the C/O ratio, which presents an upturn at low metallicity. Note that the high N/O and the C/O upturn of the low-metallicity stars is also observed in low-metallicity DLAs \citep{pettini07}.

Their CEM is computed with the assumption that yields from $Z=10^{-8}$ stellar models are valid down to $Z=0$. But actually, there is a physical transition at $Z_\mathrm{trans} \simeq 10^{-10}$: the stars below $Z_\mathrm{trans}$ present a PopIII-like behaviour (with some He burning already during the MS), and the stars above $Z_\mathrm{trans}$ have well separated H- and He-burning phases.

We computed a similar CEM, but this time including our $Z=0$ pre-SN yields up to $Z_\mathrm{trans}$, the yields of \citet{hir07} up to $Z=10^{-5}$, and then the yields from \citet{mm8} and \citet{hmm12}. For the PopIII stellar models, we used an IMF that is truncated at 1 $M_{\sun}$ but otherwise standard. Compared to the chemical evolution computed using the $Z=10^{-8}$ yields down to $Z=0$ (Fig.~\ref{fchemev}, \emph{left panel}), the new computation predicts a reduction of the N/O and C/O ratios at extremely low metallicity ($Z\leq 6\cdot 10^{-7}$). This trend is expected given the higher O and the lower C and N production of the PopIII models (Sect.~\ref{syields} and Fig.~\ref{fycomp}). Note that a similar CEM computed with the ``wind-only'' contribution for our 25 to 60 $M_{\sun}$ models leads to such a slow O enrichment that it would deviate from the CEM with $Z=10^{-8}$ yields curve only at a O/H domain much lower than that shown in Fig.~\ref{fchemev}.

Another interesting quantity is the ratio $^{12}$C/$^{13}$C. A large N production comes with a high $^{13}$C value since they are both the products of $^{12}$C conversion by the CN-cycle in H-burning zones. We thus expect a low $^{12}$C/$^{13}$C ratio from the yields of fast-rotating models. \citet{chiap08} use the same set of yields as in \citet{chiap06}, but this time to study the $^{12}$C/$^{13}$C ratio. They show that this ratio is indeed much closer to the recent observations provided by \citet{S06} upon the inclusion of fast-rotating $Z=10^{-8}$ models. With the inclusion of our pre-SN PopIII yields (Fig.~\ref{fchemev}, \emph{right panel}), we predict that the $^{12}$C/$^{13}$C ratio should rise again slightly with decreasing [Fe/H] at extremely low metallicity.
%==============================================================================
\section{Discussion and conclusion \label{sdiscu}}

We computed a set of metal-free stellar models with masses between 9 and 200 $M_{\sun}$, following their evolution until the pre-supernova stage. The set was composed of seven differentially rotating models with initial velocity 800 km s$^{-1}$ (except for the 9 $M_{\sun}$ model which started the main sequence with 500 km s$^{-1}$), and a comparison set of non-rotating models. We showed that in PopIII stars, although the effects of rotation are important, they are less strong than in stars with even a tiny amount of metals ($Z=10^{-8}$). There are two main causes for this:
\begin{enumerate}
  \item the meridional circulation is extremely weak at $Z=0$. The dependance of the outer cell in $1/\rho$ (see Sect.~\ref{sveq}) makes it a factor 100 lower compared to the value at standard metallicity, and still a factor 4 compared to $Z=10^{-5}$. The evolution of the angular momentum is close to local conservation, so the critical limit is reached late. Moreover, only a very small amount of mass needs to be lost in order to bring the surface back to sub-critical rotation;
  \item there is little structural readjustment at the end of the MS, because the core is already hot enough to burn some He during the main sequence, so the nuclear burnings are continuous. The transition between H and He burning is smooth and the models remain on the blue side of the Hertzsprung-Russell diagram for a long part of the core He burning. When they eventually reach the red side, their outer convective zone remains thin, so the dredge-up is not efficient and the surface enrichment stays low. The envelope remains transparent to radiation and the mass loss is negligible.
\end{enumerate}
There is actually a mass loss where none was expected (because of the wind dependance on $Z$), but it remains modest even for the most massive of our models.

Could this picture be modified by changing the physical inputs in the models? We have shown that the situation is qualitatively the same when we take a faster initial rotation rate. The main problem is the weakness of the meridional circulation, which does not efficiently feed the outer layers with angular momentum taken from the fast spinning core. So, even if the critical limit is reached a little earlier, the models still do not lose a lot of mass. For this work, we did not take into account the anisotropy of the winds in our models. We do not think it would change the picture presented here, because the loss of angular momentum by radiative mass loss is negligible. There is still an interesting possibility: the implementation of magnetic fields. They have been shown to change many features in stellar evolution \citep{hws05,mmB3}, and to provide the strong core-envelope coupling that is lacking in the $Z=0$ models. With such a coupling, the mass loss becomes much larger and the evolution may be strongly altered \citep{sekauai07}.

With the initial velocities considered here, our models do not follow a chemically homogeneous evolution \citep{Maed87,heglang00}. In the frame of the physics we used, homogeneous evolution for Pop III stars would need faster rotation rates to occur. When the magnetic dynamo mechanism proposed by \citet{Spruit99,Spruit02} is taken into account, chemical mixing becomes more efficient and homogeneous evolution can be obtained for smaller initial rotation rates \citep{yoon05,yoon06,woosheg06}. Note however that for extremely low metallicities, the weakening of the meridional circulation becomes a real handicap for the models to evolve chemically homogeneously. \citet{hir08} find that at $Z=10^{-8}$, a 40 $M_{\sun}$ model (computed with magnetic fields) starting its evolution with $\upsilon/\upsilon_{\rm crit}=0.55$ does not rotate fast enough to follow such an homogeneous evolution. We expect that at $Z=0$ the problem would be still more acute. The consequences of such extreme effects of rotation is left for a future work.

It is important to underline here that the only prescription for stellar dynamo existing to date, the Tayler-Spruit formalism \citep{Spruit99,Spruit02}, is still debated:  recent works have shed some doubts on the efficiency of the Tayler-Spruit dynamo mechanism. For instance, \citet{zbm07} have shown that although the Pitts and Tayler instability indeed develop, it does not succeed in driving a dynamo. Moreover, this dynamo needs a pre-existing magnetic-field seed, and it is not yet clear whether magnetic fields were present in the early Universe or not \citep[see][for a detailed discussion on this subject]{giov04}.

We presented pre-supernova yields for our stellar models. We included them in a chemical evolution model of the Galactic halo. Our results predict that the N/O and the C/O ratios should drop at extremely low metallicity ($Z\leq 10^{-6}$). For the time being though, the observational discrimination between the two curves is far beyond our observational possibilities. We also gave the rotational characteristics of the iron core of our models and showed that since they do not lose much angular momentum during their evolution, they keep high rotation rates in the core, so their final explosion may be much more energetic than sometimes assumed. We showed that the most massive of our models may contribute to the emission of the gravitational wave background.

Many interesting questions and issues were not treated in this study. Our mass coverage was too coarse to constrain accurately the limiting masses for undergoing a CNO shell boost, for example, or for the different fates at the end of the evolution. Even without knowing if magnetic fields were indeed present in the early Universe, it would be interesting to study the role they could play in primordial stellar evolution. As mentioned above, they could drastically change the picture presented here. We have also left for a forthcoming paper the question of the re-ionisation. Rotation changes the luminosity, the $T_\mathrm{eff}$ and the lifetimes of the models, so it should change the ionising power of the first stars. 
%==============================================================================
\bibliographystyle{aa}
\bibliography{aa9633-08}

\end{document}